\documentclass[12pt]{article}

\pdfoutput=1
\usepackage{latexsym}
\usepackage{epsfig,amssymb,euscript,slashed}
\usepackage{amsmath}
\usepackage{cite}
\usepackage{array,calc,epsfig}
\usepackage[linktocpage]{hyperref}
\usepackage{pstricks}
\usepackage{bbm}
\usepackage{fancybox}

\def\be{\begin{equation}}
\def\ee{\end{equation}}
\def\bseq{\begin{subequations}}
\def\eseq{\end{subequations}}

\def\bea{\begin{eqnarray}}
\def\eea{\end{eqnarray}}
\newcommand\bbone{\ensuremath{\mathbbm{1}}}

\def\bseq{\begin{subequations}}
\def\eseq{\end{subequations}}

\numberwithin{equation}{section} 
\usepackage[]{graphicx}

\def\d {{\rm d}}

\def\kahler                        {{K\"ahler\ }} 
\def\cala         {{\cal A}}

\def\calb         {{\cal B}}
\def\calc         {{\cal C}}
\def\cald         {{\cal D}}

\def\calf         {{\cal F}}

\def\calh         {{\cal H}}

\def\call         {{L}}

\def\calo         {{\cal O}}

\def\calq         {{\cal Q}}
\def\calr         {{\cal R}}

\def\calv         {{\cal V}}

\def\del          {\partial}
\def\delbar       {\bar\partial}
\def\ii           {{\rm i}}

\def\Re           {{\rm Re\hskip0.1em}}
\def\Im           {{\rm Im\hskip0.1em}}

\def\sqr#1#2{{\vcenter{\vbox{\hrule height.#2pt
 \hbox{\vrule width.#2pt height#1pt \kern#1pt \vrule width.#2pt}\hrule
 height.#2pt}}}}

\def\d{\text{d}}

\def\slashchar#1{\setbox0=\hbox{$#1$}           
\dimen0=\wd0                                 
\setbox1=\hbox{/} \dimen1=\wd1               
\ifdim\dimen0>\dimen1                        
\rlap{\hbox to \dimen0{\hfil/\hfil}}      
#1                                        
\else                                        
\rlap{\hbox to \dimen1{\hfil$#1$\hfil}}   
/                                         
\fi}

\begin{document}
\font\cmss=cmss10 \font\cmsss=cmss10 at 7pt

\vskip -0.5cm
\rightline{\small{\tt DFPD-14/TH/03}}

\vskip .7 cm

\hfill
\vspace{18pt}
\begin{center}
{\Large \textbf{Topological Duality Twist}}
 
\smallskip

{\Large \textbf{and Brane Instantons in F-theory}}
\end{center}

\vspace{6pt}
\begin{center}
{\large\textsl{ Luca Martucci}}

\vspace{25pt}
\textit{\small Dipartimento di Fisica ed Astronomia ``Galileo Galilei",  Universit\`a  di Padova\\
\& I.N.F.N. Sezione di Padova,
Via Marzolo 8, I-35131 Padova, Italy} 
\end{center}

\vspace{12pt}

\begin{center}
\textbf{Abstract}
\end{center}

\vspace{4pt} {\small
\noindent A variant of the topological twist, involving ${\rm SL}(2,\mathbb{Z})$ dualities and hence named topological duality twist, is introduced and explicitly applied to describe a  U(1) $N=4$ super Yang-Mills theory on a \kahler space  with holomorphically space-dependent coupling. Three-dimensional duality walls
and two-dimensional chiral theories naturally enter the formulation of the duality twisted theory. Appropriately generalized, this theory is relevant for the study of Euclidean D3-brane instantons  in F-theory compactifications. Some of its properties and implications are discussed.

\noindent }

\vspace{1cm}


\thispagestyle{empty}

\vfill
\vskip 5.mm
\hrule width 5.cm
\vskip 2.mm
{\small
\noindent e-mail: luca.martucci@pd.infn.it
}

\newpage

\setcounter{footnote}{0}


\vspace{0.5cm}


\section{Introduction}

There is strong evidence  that $N=4$ Super Yang-Mills (SYM)  in four dimensions with gauge group U($n$) 
is self-dual under the duality group $\text{SL(2,}\mathbb{Z})$, see for instance \cite{Harvey:1996ur,DiVecchia:1998ky} for introductions to the subject. In particular, under an element 
\be\label{gamma}
\gamma=\left(\begin{array}{cc}a & b \\ c & d\end{array}\right)\in \text{SL(2,}\mathbb{Z})
\ee
the complexified coupling constant
\be
\tau\equiv \frac{\theta}{2\pi}+\frac{4\pi\ii}{g_{\rm YM}^2}
\ee
transforms as  
\be\label{taudual}
\tau\rightarrow \tau'=\gamma\cdot\tau\equiv \frac{a\tau+b}{c\tau+d}
\ee
One usually takes $\tau$ to be constant along the four-dimensional space-time. 

This paper will instead  consider a specific class of deformations of $N=4$ SYM in which the coupling $\tau$ is not constant and can undergo non-trivial $\text{SL(2,}\mathbb{Z})$ monodromies, but which still preserve a certain amount of supersymmetry.\footnote{See for instance \cite{D'Hoker:2006uv,Harvey:2007ab,Harvey:2008zz,Gaiotto:2008sd,Ganor:2010md,Ganor:2014pha} for other constructions of this kind.} The mechanism we will use to construct such models  is analogous to the one originally introduced in \cite{Witten:1988ze}. In that case, one can define a SYM theory
on a curved space  by going to Euclidean formulation and performing a topological twist,  which basically allows some combination of the supercharges to be globally defined over the curved space. We apply a similar trick to construct supersymmetric theories with non-constant $\tau$ and $\text{SL(2,}\mathbb{Z})$ dualities. We dub such procedure {\em topological duality twist} (TDT).

 More precisely, we explore this possibility in a case which is relatively simple.  First, we focus on the case with abelian U(1) gauge group, for which the $\text{SL(2,}\mathbb{Z})$ duality is well understood. As we will see,   even if the abelian case is substantially simpler than the non-abelian one, the resulting theory will be non-trivial enough to show some interesting properties which should be shared by the non-abelian case as well. 
Furthermore, we put the $N=4$ SYM theory on a K\"ahler space, allowing $\tau$ to depend holomorphically on the complex coordinates, $\tau=\tau(z)$. 
 In fact, such choices have as concrete physical motivation the study of supersymmetric (Euclidean) D3-brane instantons in F-theory compactifications started in \cite{Witten:1996bn}, see \cite{Cvetic:2009ah,Blumenhagen:2010ja,Cvetic:2010rq,Donagi:2010pd,Cvetic:2010ky,Grimm:2011dj,Marsano:2011nn,Bianchi:2011qh,Bianchi:2012pn,Kerstan:2012cy,Cvetic:2012ts,Bianchi:2012kt} for recent work on the subject. 
Indeed, by supersymmetry the D3-brane must wrap a K\"ahler submanifold of the compactification space and the  coupling $\tau$ just corresponds to the type IIB axio-dilaton, whose non-trivial holomorphic profile characterizes the F-theory vacua. Furthermore, by extrapolating the arguments of \cite{Bershadsky:1995qy} to the F-theory context, one expects the brane theory to be somehow topologically twisted, see for instance  \cite{Heckman:2008es}. As we will discuss, one should actually  implement a topological duality twist.

 In such a setting   $\tau(z)$ experiences non-trivial monodromies as one moves around and two-dimensional defects and duality walls naturally enter the story. 
  In particular, the resulting theory combines the four-dimensional SYM theory with Chern-Simons like terms supported on duality walls \cite{Ganor:1996pe,Gaiotto:2008ak,Kapustin:2009av} and two-dimensional  chiral theories living on defects, hence named {\em chiral defects}, along which the duality walls can end.  
 Such peculiar four-dimensional  theories are directly related to the  (2,0) six-dimensional theory,  as they can be obtained by compactifying the latter on an elliptically fibered  K\"ahler space and sending the size of the elliptic fiber  to zero. 
The complex structure of the elliptic fiber then corresponds to the four-dimensional coupling $\tau(z)$ and the $\text{SL(2,}\mathbb{Z})$ duality group can be identified with the modular group of the elliptic fiber \cite{Verlinde:1995mz}. This parent six-dimensional theory
describes, in a certain approximation, the (Euclidean) M5-branes  wrapping `vertical' divisors of the  elliptically fibered Calabi-Yau four-fold which is associated with the  F-theory compactification. The non-trivial structure of the duality twisted four-dimensional theory hence encodes  the  subtleties of the chiral two-form theory 
supported on the M5-brane. From this viewpoint, the topological duality twist of the $N=4$ SYM theory may open a new perspective over the study of the still elusive $(2,0)$ theory describing M5-branes.\footnote{This is in fact the strategy already followed, at the bosonic level, in \cite{Ganor:1996pe} in order 
to understand from a purely type IIB perspective  the results of  \cite{Witten:1996hc}  on the axionic moduli dependence of the M5-brane partition function.} 
This paper will mostly concern the classical structure of the duality twisted theories. Other important questions regarding the quantization of the theory
as well as the application to F-theory compactifications will be  discussed elsewhere.

The paper is organized as follows. Section \ref{sec:group} describes the group-theoretical structure of the TDT in quite general terms.
In section \ref{sec:TDT} it is shown how to explicitly realize the TDT in the case of U(1) gauge group. Section \ref{sec:dwc} concerns 
one of the distinguishing features of the duality twisted theory: the presence of duality walls \cite{Ganor:1996pe,Gaiotto:2008ak,Kapustin:2009av} and chiral defects.  Section \ref{sec:6d} reviews some basic results on elliptically fibered three-folds, which are useful for describing the global structure of the duality twisted SYM theory, presenting  a concrete example on a Hirzebruch surface. Section \ref{sec:Ftheory} discusses more explicitly how the results developed in the previous sections can be embedded in the context of F-theory compactifications, in particular by including the dependence on the bulk axionic moduli. As an  application, in section \ref{sec:application} it is shown how the duality twisted structure can be used to  prove that the partition function of the theory does not depend on the bulk \kahler structure.
Section \ref{sec:concl} contains some concluding remarks.


\section{Group theoretical structure of the TDT}
\label{sec:group}

We start by describing some general aspects of the  topological duality twist  for $N=4$ SYM on a K\"ahler space.
Although we will  explicitly realize it  only in the drastically simpler case of U(1) gauge group, several aspects of  the discussion of this section are more general 
and should apply to more general gauge groups.   

Let us first recall   few basic facts about  $N=4$ SYM theory.
The field content is constituted by a gauge field $A_m\sim A_{A\dot B}$ taking values in the gauge algebra, six scalars $\varphi^{i}$ , and eight fermions $\Psi^I_{A}$ and $\tilde\Psi_{I\dot A}$. The indices $m,n,\ldots=1,\ldots,4$ denote the (Euclidean) space indices, $A=1,2$ ($\dot A=\dot 1,\dot 2$) denote the left-(right-)moving space Weyl indices ,  $I=1,\ldots,4$ and $i=1,\ldots,6$ transform in the ${\bf 4}$ or $\bar{\bf 4}$  and ${\bf 6}_{\bf v}$ respectively  of the internal $\calr$-symmetry group ${\rm SU(4)}_{\calr}\simeq {\rm Spin}(6)_\calr$.  Namely, by writing the  space rotation  group as ${\rm Spin(4)}\simeq{\rm SU(2)}_L\times{\rm SU(2)}_R$, under ${\rm SU(2)}_L\times{\rm SU(2)}_R\times {\rm SU(4)}_{\calr}$ the fields transform as
\be
A_\mu \in({\bf 2},{\bf 2},{\bf 1}) \quad~~~~ \varphi^{i}\in ({\bf 1},{\bf 1},{\bf 6}_{\bf v}) \quad~~~~ \Psi^I_{A}\in({\bf 2},{\bf 1},{\bf 4})\quad~~~~ \tilde\Psi_{I\dot A}\in ({\bf 1},{\bf 2},\bar{\bf 4})
\ee
 On the other hand, the sixteen supercharges transform as 
 \be
 Q_{A I}\in ({\bf 2},{\bf 1},\bar{\bf 4}) \quad~~~~~~~ \tilde Q_{\dot A}^I\in ({\bf 1},{\bf 2},{\bf 4})
 \ee
Notice that in Euclidean space, differently from the Minkowskian case,  the elements of the pairs of fermions $(\Psi^I_{A},\tilde\Psi_{I\dot A})$ and $(Q_{A I},{\tilde Q}_{\dot A}^I)$ 
 are not related by complex conjugation.  

We will also need the transformation properties of the supercharges under  $\gamma\in {\rm SL(2,\mathbb{Z})}$ as in  (\ref{gamma}) \cite{Kapustin:2006pk}. In order to describe them, let us associate an element $\gamma\in {\rm SL(2,\mathbb{Z})}$ to
a phase   $e^{\ii\alpha(\gamma)}\in {\rm U(1)}_\cald$ defined by  
\be
e^{\ii\alpha(\gamma)}=\frac{c\tau+d}{|c\tau+d|}
\ee  
We then say that an object has ${\rm U(1)}_\cald$-charge $q_\cald$ if it transforms by a phase $e^{\ii q_\cald\alpha(\gamma)}$ under the duality $\gamma$. It turns out that 
$Q_{A I}$  and ${\tilde Q}_{\dot A}^I$ have  $q_\cald$-charges $+\frac12$ and $-\frac12$ respectively, namely
\be\label{Qduality}
Q_{A I}\rightarrow e^{\frac{\ii}{2}\alpha(\gamma)}Q_{A I}\quad~~~~~~~~~~~
{\tilde Q}_{\dot A}^I\rightarrow e^{-\frac{\ii}{2}\alpha(\gamma)}{\tilde Q}^I_{{\dot A}}
\ee
Analogously, the pair 
$(\Psi^I_{A},\tilde\Psi_{I\dot A})$ 
 transforms with  ${\rm U(1)}_\cald$-charges $(+\frac12,-\frac12)$.
These simple transformation rules will play a crucial role in the following. 

A  point-dependent coupling $\tau$ can be used to construct a composite connection  $\cala$ for the group ${\rm U(1)}_\cald$ in a way familiar from type IIB supergravity:
 \be\label{conn}
 \cala=\frac{1}{2\Im\tau}\,\d\Re\tau
 \ee  
Hence, for non-constant $\tau$, $\cala$ defines a ${\rm U(1)}_\cald$ line bundle $\call_\cald$ and  the fields which transform with ${\rm U(1)}_\cald$-charge $q_\cald$ can be regarded as sections of $\call_\cald^{q_\cald}$.

Coupling the theory to a curved space generically breaks supersymmetry since the non-trivial holonomy makes the supercharges not well defined anymore. The strategy of the standard topological twist \cite{Witten:1988ze} -- see \cite{Yamron:1988qc,Vafa:1994tf,Dijkgraaf:1997ce} for discussions on the  $N=4$ SYM case -- is to accompany  the non-trivial holonomy under the Lorentz group  by a non-trivial holonomy under the  $\calr$-symmetry group, so that one or more supercharges are singlets under the combined action and can still be globally defined. 
We want to generalize this strategy to include a non-constant $\tau$ which can undergo non-trivial ${\rm SL(2,\mathbb{Z})}$ dualities as we move around.
Since the supercharges transform non-trivially  under ${\rm SL(2,\mathbb{Z})}$,  generically they fail to be globally well defined. 
On the other hand, they transform just as half-charged objects under ${\rm U(1)}_\cald$,  as in (\ref{Qduality}). Hence, 
one can apply the same trick as for the standard topological twist by compensating the non-trivial ${\rm U(1)}_\cald$  transformations by a corresponding $\calr$-symmetry transformation. 

We will focus on a specific realization of this general idea, which is the relevant one for applications to brane instantons in F-theory compactifications. 
Namely, we put the theory on a K\"ahler manifold $S$, with K\"ahler form $j$. The holonomy group is then restricted to  ${\rm SU(2)}_L\times{\rm U(1)}_J$, 
where ${\rm U(1)}_J\subset {\rm SU(2)}_R$ rotates locally flat complex coordinates by a phase. Under ${\rm U(1)}_J$ the two components $\eta_{\dot 1}$ and $\eta_{\dot 2}$  of a right-handed spinor $\eta_{\dot A}$ transform with charges $+1$ and $-1$ respectively. Furthermore, we split ${\rm SU(4)}_{\calr}\simeq {\rm Spin(6)}_{\calr}$ into 
${\rm Spin(4)}_{\calr}\times {\rm U(1)}_{\calr}\simeq {\rm SU(2)}_{A}\times {\rm SU(2)}_{B}\times{\rm U(1)}_{\calr}$
and we require $\tau$ to be holomorphic: 
\be
\delbar\tau=0
\ee
Then the ${\rm U(1)}_\cald$ connection $\cala$ defines a holomorphic line bundle, $F^{0,2}_\cala=\delbar\cala^{0,1}=0$, and 
we can split the covariant exterior derivative $\d_\cala=\d-\ii q_\cald\cala$
 into holomorphic and anti-holomorphic parts
\be
\d_\cala=\del_\cala+\delbar_\cala\quad~~~~~~~~~\text{with}\quad \del_\cala^2=\delbar_\cala^2=0
\ee

Under these conditions, the relevant group is 
\be
G={\rm SU(2)}_L\times{\rm SU(2)}_{A}\times {\rm SU(2)}_{B}\times {\rm U(1)}_J\times{\rm U(1)}_{\calr}\times {\rm U(1)}_\cald
\ee
Accordingly, the supercharges split
into the following reduced representations:
\be\label{Qdec}
\begin{aligned}
Q_{A I}&\rightarrow ({\bf 2},{\bf 2},{\bf 1})_{0,\frac12,\frac12}\oplus ({\bf 2},{\bf 1},{\bf 2})_{0,-\frac12,\frac12} \\
\tilde Q^I_{\dot A}&\rightarrow ({\bf 1},{\bf 2},{\bf 1})_{1,-\frac12,-\frac12}\oplus ({\bf 1},{\bf 2},{\bf 1})_{-1,-\frac12,-\frac12}\oplus({\bf 1},{\bf 1},{\bf 2})_{1,\frac12,-\frac12}\oplus({\bf 1},{\bf 1},{\bf 2})_{-1,\frac12,-\frac12}
\end{aligned}
\ee
where the triplet in brackets  gives the representations under ${\rm SU(2)}_L\times{\rm SU(2)}_A\times {\rm SU(2)}_B$ and the subscripts indicate the charges under $ {\rm U(1)}_J\times{\rm U(1)}_{\calr}\times {\rm U(1)}_\cald$. 

We look for a twisted theory in which  ${\rm SU(2)}_L\times {\rm U(1)}_J\times {\rm U(1)}_\cald$ can have non-trivial holonomies by twisting them with ${\rm U(1)}_{\calr}$, while ${\rm Spin(4)}_{\calr}\simeq{\rm SU(2)}_{A}\times {\rm SU(2)}_{B}$ survives as an external rigid symmetry group. 
Let us denote by $J$, $\calr$ and $\cald$ the generators of  $ {\rm U(1)}_J$, ${\rm U(1)}_{\calr}$ and ${\rm U(1)}_\cald$ respectively.
In the duality twisted theory these groups are substituted by the twisted groups $ {\rm U(1)}'_J$ and  ${\rm U(1)}'_\cald$ associated 
with the generators\footnote{We could equivalently choose different relative signs in defining the twisted generators.}
\be\label{dtw}
J'\equiv J+ 2\calr\quad~~~~~~~~~~~~ \cald'=\cald+\calr
\ee
Then, it is immediate to check that the $({\bf 1},{\bf 1},{\bf 2})_{-1,\frac12,-\frac12}$ component of $\tilde Q^I_{\dot A}$ provides two supercharges $\tilde\calq_{\dot\alpha}$ which transform as
\be
\tilde\calq_{\dot\alpha}\in ({\bf 1},{\bf 1},{\bf 2})_{0,0}
\ee
under 
\be\label{classgroup1}
G'={\rm SU(2)}_L\times{\rm SU(2)}_{A}\times {\rm SU(2)}_{B}\times {\rm U(1)}'_J\times {\rm U(1)}'_\cald
\ee where we have used Weyl indices $(\alpha,\dot\beta)$ for 
 ${\rm Spin(4)}_\calr\simeq {\rm SU(2)}_{A}\times {\rm SU(2)}_{B}$.  Namely the supercharges $\tilde\calq_{\dot\alpha}$ are singlets under ${\rm SU(2)}_L\times {\rm U(1)}'_J\times {\rm U(1)}'_\cald$ for which we allow non-trivial holonomies, while they transform as right-handed spinors  under the surviving global symmetry  group ${\rm Spin(4)}_\calr$. Hence,
 we say that the resulting theory have {\em chiral} $(0,2)$ twisted supersymmetry, keeping in mind that we classifying the twisted supersymmetry according to the external symmetry  group ${\rm Spin(4)}_\calr$.

Notice that our topological duality twist is very similar to the ordinary topological twist which does not involve $\cald$ at all and just twists $J$ into $J'\equiv J+ 2\calr$.\footnote{See for instance  \cite{Heckman:2008es} for a discussion on this topological twist  in in the context of F-theory compactifications.} In this case  there are two additional twisted supersymmetries $\calq_\alpha$  arising from the relabelling of the $({\bf 1},{\bf 2},{\bf 1})_{+1,-\frac12,-\frac12}$ component of $\tilde Q^I_{\dot A}$ in (\ref{Qdec}). In that case the resulting theory has {\em non-chiral} $(2,2)$ topologically twisted supersymmetry
and we see how one of the effects of including the duality generator $\cald$ in the twist is to make it  chiral.


\section{TDT of  U(1)  $N=4$ SYM: local structure}
\label{sec:TDT}

We now restricts to an $N=4$ theory with U(1) gauge group,
for which the  ${\rm SL(2,\mathbb{Z})}$ duality is well understood.
Of course, compared to the more general non-abelian case, this is a drastically simpler setting. However , as we will see, even in this case the TDT presents several non-trivial features. 

In flat space, the (Euclidean) U(1) $N=4$ SYM action  is given by 
\be\label{flataction}
\begin{aligned}
I_{\rm SYM}&=\int_{\mathbb{R}^4} \Big(\frac1{g^2_{\rm YM}}*F\wedge F-\frac{\ii\theta}{8\pi^2} F\wedge F\Big)+\frac1{4\pi} \int_{\mathbb{R}^4}*\d\varphi^i\wedge \d\varphi^i
+\frac{\ii}{2\pi}\int_{\mathbb{R}^4}\d^4 x \,\tilde\Psi_I\bar\sigma^m\del_m\Psi^I
\end{aligned}
\ee
In order to proceed with its TDT, we have first to understand how the different fields decompose according to twisted classifying group (\ref{classgroup1}).

 We start form  the scalar fields $\varphi^{i}$, which
transform as ${\bf 6_{v}}\simeq ({\bf 4}\otimes {\bf 4})_{\rm A}$ of ${\rm Spin(6)}_{\calr}\simeq {\rm SU(4)}_{\calr}$.
Hence, by splitting ${\rm Spin(6)}_{\calr}\rightarrow  {\rm SU(2)}_{A}\times {\rm SU(2)}_{B}\times{\rm U(1)}_{\calr}$ the scalar fields decompose into four real scalar fields $\varphi^\mu\sim \varphi^{\alpha\dot\beta}$, $\mu=1,\ldots, 4$, and a pair of complex fields $\sigma$ and $\tilde\sigma$ (not to be confused with the Weyl matrices $\sigma^m_{A\dot B}$, $\bar\sigma^{\dot A B}_m$ and $\sigma^\mu_{\alpha\dot \beta}$, $\bar\sigma^{\dot \alpha \beta}_\mu$). Before the TDT, they are neutral under ${\rm SL}(2,\mathbb{Z})$ dualities. Hence,
after the TDT,  the  scalars $\varphi^i$ are reorganized as follows under  (\ref{classgroup1}) 
\be
\begin{aligned}
\varphi^i\rightarrow\{\varphi^{\alpha\dot\beta}\in ({\bf 1},{\bf 2}, {\bf 2})_{0,0}\}\oplus\{
\sigma\in ({\bf 1 },{\bf 1}, {\bf 1})_{2,1}\}\oplus\{
\tilde\sigma\in ({\bf 1},{\bf 1}, {\bf 1})_{-2,-1}\}
\end{aligned}
\ee
We can then identify $\sigma$ and $\tilde\sigma$ 
with a $(2,0)$ form and a $(0,2)$ form taking values in $\call_\cald$ and $\call_\cald^{-1}$ respectively:
\be
\sigma\equiv\frac12\,\sigma_{ab}\,\d s^a\wedge \d s^b\quad~~~~~~~~~~~\tilde\sigma\equiv\frac12\,\tilde\sigma_{\bar a\bar b}\,\d \bar s^{\bar a}\wedge \d \bar s^{\bar b}
\ee
where we have introduced complex coordinates $s^a$, $a=1,2$, on $S$. 

Let us now consider the fermions. Under the twisted group (\ref{classgroup1})  they decompose as
\be
\begin{aligned}
\Psi^I_{A}&\rightarrow\{ \psi^\alpha\in ({\bf 2},{\bf 2},{\bf 1})_{-1,0}\}\oplus \{ \tilde\psi^{\dot \alpha}\in ({\bf 2},{\bf 1},{\bf 2})_{1,1}\} \\
\tilde\Psi_{I\dot A}&\rightarrow \{\rho^\alpha\in({\bf 1},{\bf 2},{\bf 1})_{2,0}\}\oplus \{\lambda^\alpha\in ({\bf 1},{\bf 2},{\bf 1})_{0,0}\}
\\ &\quad~~~~~~~~~~~~~\oplus\{\tilde\lambda^{\dot\alpha}\in({\bf 1},{\bf 1},{\bf 2})_{0,-1}\}\oplus\{\tilde\rho^{\dot\alpha}\in({\bf 1},{\bf 1},{\bf 2})_{-2,-1}\}
\end{aligned}
\ee
Hence,  fermions transform as forms (as in standard topologically twisted theories), which in addition take values in some power of $\call_\cald$. Namely, $\lambda^\alpha$ and $\tilde\lambda^{\dot\alpha}$ transform as scalars under space rotation, while
\be
\begin{aligned}
\psi^\alpha&\equiv\psi^\alpha_{\bar a}\d \bar s^{\bar a}
\quad~~~~~~~~~~~~~~~~~~~~~~~~\tilde\psi^{\dot\alpha}\equiv\tilde\psi^{\dot\alpha}_a\d s^a\\
\rho^\alpha&\equiv\frac12\,\rho^\alpha_{ ab}\,\d  s^{a}\wedge \d s^b\quad~~~~~~~~~~~~~\tilde\rho^{\dot\alpha}
\equiv\frac12\,\tilde\rho^{\dot\alpha}_{\bar a\bar b}\,\d \bar s^{\bar a}\wedge \d\bar s^{\bar b}
\end{aligned}
\ee

It remains to discuss the gauge field. It transforms non-trivially under the ${\rm SL}(2,\mathbb{Z})$ duality transformation.
 In general, the dual field-strength is given by
\be\label{dualF}
F_{\rm D}=2\pi\ii\,\frac{\delta I_{\rm SYM}}{\delta F}
\ee
where  $\frac{\delta I_{\rm SYM}}{\delta F}$ is defined by $\delta I_{\rm SYM}=\int_S\delta F\wedge \frac{\delta I_{\rm SYM}}{\delta F}$.
Under an ${\rm SL}(2,\mathbb{Z})$ duality (\ref{gamma}) we have
\be\label{dualityF}
\left(\begin{array}{c} F_{\rm D}  \\  F\end{array}\right)\rightarrow \left(\begin{array}{cc}a & b \\ c & d\end{array}\right)\left(\begin{array}{c} F_{\rm D} \\ F\end{array}\right)
\ee
See for instance \cite{Witten:1995gf} for a path-integral derivation of this duality.

One can decompose the field-strength in components of definite   ${\rm U(1)}_\cald$ charge:
\be\label{sdfluxes}
\begin{aligned}
\sqrt{\Im\tau}\,F_+&\equiv-\frac{\ii}{2\sqrt{\Im\tau}}(F_{\rm D}-\bar\tau F) \quad~~~~~~~~~ (q_\cald=+1)\\
\sqrt{\Im\tau}\,F_-&\equiv\frac{\ii}{2\sqrt{\Im\tau}}(F_{\rm D}-\tau F) \quad~~~~~~~~~~~ (q_\cald=-1)
\end{aligned}
\ee
One can easily compute $F_{\rm D}=\frac{\theta}{2\pi}F+\frac{4\pi\ii}{g^2_{\rm YM}}*F$ and then
\be
F_\pm=\frac12(1\pm *)F
\ee 
Since the gauge field is not charged under the $\calr$-symmetry group, the world-volume gauge field is not affected by the TDT and so preserves its nature.  Furthermore, given the K\"ahler structure on $S$, we can  identify $F_+$ with the (primitive)  $(1,1)_{\rm P}$  component of $F$ and $F_-$ with its $(2,0)$, $(0,2)$ and (non-primitive) $(1,1)_{\rm NP}$ components.

To summarize, we have arrived at the following spectrum of duality twisted fields
\begin{equation}\label{fcharges}
\begin{tabular}{ | c | c | }
  \hline 
    boson &
     bundle \\
       \hline  
       $\sqrt{\Im\tau}\,F_+ $ & $\Lambda^{1,1}_{\rm P}\otimes \call_\cald$   \\
        $\sqrt{\Im\tau}\,F_- $  & $(\Lambda^{2,0}\oplus\Lambda^{0,2}\oplus \Lambda^{1,1}_{\rm NP})\otimes \call^{-1}_\cald$  \\
        $\varphi^{\alpha\dot\beta} $  & $\calo_S$  \\
        $\sigma$  & $\Lambda^{2,0}\otimes \call_\cald$ \\
        $\tilde\sigma $  & $\Lambda^{0,2}\otimes \call^{-1}_\cald$  \\
    \hline  
\end{tabular}
\quad~~~~
\begin{tabular}{ | c | c | }
  \hline 
   fermion &
    bundle \\
     \hline  
  $\lambda^\alpha$ & $\calo_S$  \\
   $ \psi^\alpha$ & $\Lambda^{0,1}$  \\
    $\rho^\alpha$ & $\Lambda^{2,0}$   \\
   $\tilde\lambda^{\dot\alpha}$  & $\calo_S\otimes \call^{-1}_\cald$  \\
  $\tilde\psi^{\dot\alpha}$& $\Lambda^{1,0}\otimes \call_\cald$ \\
   $\tilde\rho^{\dot\alpha}$  &  $\Lambda^{0,2}\otimes \call^{-1}_\cald$    \\
    \hline  
\end{tabular}
\end{equation}
Here $\calo_S$ denotes the trivial line bundle on $S$, $\Lambda^{p,q}$ refers to the bundle of $(p,q)$ forms on $S$ and the subscript $_{\rm P}$ and $_{\rm NP}$ refers to the primitive and non-primitive component respectively.

We are now ready to write down the duality twisted theory. The flat space action (\ref{flataction}) is replaced by the following four-dimensional action
\be\label{dtwaction}
\begin{aligned}
I_{\rm 4d}&=I_{\rm YM}+I_{\rm B}+I_{\rm F}\\
I_{\rm YM}&=\int_S \Big(\frac1{g^2_{\rm YM}}*F\wedge F-\frac{\ii\theta}{8\pi^2} F\wedge F\Big)=
-\frac{\ii}{4\pi}\int_S\tau F\wedge F  - \frac{1}{2\pi}\int_S\Im\tau F_-\wedge F_-\\
 I_{\rm B}&=-\frac{\ii}{4\pi}\int_S\Big( j\wedge \del\varphi^{\alpha\dot\beta}\wedge \delbar\varphi_{\alpha\dot\beta}  -2 j\wedge \del_\cala^\dagger\sigma\wedge\delbar_\cala^\dagger\tilde\sigma\Big)\\
I_{\rm F}&=\frac{1}{\pi}\int_S\Big(j\wedge \del\lambda^\alpha\wedge\psi_\alpha-\ii\rho^\alpha\wedge \delbar\psi_\alpha+j\wedge\tilde\psi_{\dot\alpha}\wedge \delbar_\cala\tilde\lambda^{\dot\alpha}-\ii\tilde\rho_{\dot\alpha}\wedge \del_\cala\tilde\psi^{\dot\alpha}\Big)
\end{aligned}
\ee
Notice  $ I_{\rm B}$ and $I_{\rm F}$ are manifestly ${\rm SL}(2,\mathbb{Z})$ invariant, as can be checked 
by looking at the transformation properties of the fields, collected in (\ref{fcharges}). Of course, there could be possible deformations of this action, preserving the properties we are going to discuss. However, the action (\ref{dtwaction}) is the supersymmetrization of (the field-theory limit of) the fermionic action identified in \cite{Bianchi:2011qh,Bianchi:2012kt}, in the context of F-theory compactifications. Then, the action (\ref{dtwaction}) is the one naturally selected for applications to F-theory  compactifications. 

 The explicit action of duality twisted supercharges $\tilde\calq_{\dot\alpha}$ on the twisted fields is given by
\be\label{dtwsuper}
\begin{aligned}
\tilde\calq_{\dot\alpha}\, A^{1,0}&=\frac{1}{\sqrt{\Im\tau}}\, \tilde\psi_{\dot\alpha} \quad~~~~~~~~~~~~~~~~~  \tilde\calq_{\dot\alpha}\, A^{0,1}=0\\
\tilde\calq_{\dot\alpha}\,  \sigma&=0\quad~~~~~~~~~~~~~~~~~~~~~~~~~~~~  \tilde\calq_{\dot\alpha}\,  \tilde\sigma=\tilde\rho_{\dot\alpha}\\
\tilde\calq_{\dot\alpha}\,  \varphi^{\beta\dot\gamma}&=\delta^{\dot\gamma}_{\dot\alpha}\lambda^\beta \\
\tilde\calq_{\dot \alpha}\, \lambda^\beta&=0\quad~~~~~~~~~~~~~~~~~~~~~~~~~~~~ 
 \tilde\calq_{\dot\alpha}\, \tilde\lambda^{\dot\beta}=-\frac14 \sqrt{\Im\tau}\,\delta^{\dot\beta}_{\dot\alpha}\,j^{mn}\,F_{mn}\\
\tilde\calq_{\dot\alpha}\, \psi_\beta&=-\frac{\ii}{2}\delbar\varphi_{\beta\dot\alpha}\quad~~~~~~~~~~~~~~~~~~ \tilde\calq_{\dot\alpha}\, \tilde\psi^{\dot\beta}=\frac{\ii}{2}\delta^{\dot\beta}_{\dot\alpha}\del_\cala^\dagger\sigma\\
\tilde\calq_{\dot\alpha}\, \rho^{\beta}&=0\quad~~~~~~~~~~~~~~~~~~~~~~~~~~~~  \tilde\calq_{\dot\alpha}\, \tilde\rho^{\dot\beta}=-\ii\sqrt{\Im\tau}\, \delta^{\dot\beta}_{\dot\alpha}\, F^{0,2}
\end{aligned}
\ee
In fact, by direct inspection, one can  check that (\ref{dtwaction}) is invariant under (\ref{dtwsuper}), {\em up to possible boundary terms},
which arise from integrations by parts. So, if we could straightforwardly apply (\ref{dtwaction}) to the whole $S$, it would provide a satisfactory solution. But, actually, this is possible only if $\tau$ is constant. Indeed any non-constant holomorphic $\tau$ generically experiences ${\rm SL}(2,\mathbb{Z})$ monodromies. So, $\tau$  is not naturally globally defined and  the action (\ref{dtwaction}), or more precisely $I_{\rm YM}$, has a global meaning only if we introduce some three-dimensional cuts on $S$ along which $\tau$ can `jump' by  an ${\rm SL}(2,\mathbb{Z})$ transformation (\ref{gamma}). In fact we will see that the variation of  (\ref{dtwaction})  under (\ref{dtwsuper}) is non-vanishing by boundary terms localized on these branch cuts, and will be compensated by  three- and two-dimensional terms which must be added to  (\ref{dtwaction}) in order to get the complete action. 

Notice that, on-shell, the duality twisted supercharges (\ref{dtwsuper}) satisfy the desirable algebra $\{\tilde\calq_{\dot\alpha},\tilde\calq_{\dot\beta}\}=0$. In fact, the only equation of motion needed is $j\wedge \delbar_\cala\tilde\psi_{\dot\alpha}=0$,  since
the off-shell violation of these anticommutation relations comes from $\{\tilde\calq_{\dot\alpha},\tilde\calq_{\dot\beta}\}\tilde\lambda^{\dot\gamma}=-\frac12j^{mn}(\delbar_\cala\tilde\psi_{(\dot\alpha})_{mn}\delta^{\dot\gamma}_{\dot\beta)}$.
As usual, one can ameliorate the situation by introducing an 
auxiliary scalar field $\calh$ transforming as a section of $\call_\cald^{-1}$ and substituting $\tilde\calq_{\dot\alpha}\tilde\lambda^{\dot\beta}=-\frac14 \sqrt{\Im\tau}\,\delta^{\dot\beta}_{\dot\alpha}\,j^{mn}\,F_{mn}$ in (\ref{dtwsuper}) with
\be\label{aux}
\begin{aligned}
 \tilde\calq_{\dot\alpha}\, \calh&=0\\
 \tilde\calq_{\dot\alpha}\, \tilde\lambda^{\dot\beta}&=-\frac12 \,\delta^{\dot\beta}_{\dot\alpha}\,\calh
\end{aligned}
\ee
Correspondingly, one has to modify $I_{\rm YM}$ in (\ref{dtwaction}) into
\be
I'_{\rm YM}=-\frac{\ii}{4\pi}\int_S\tau F\wedge F  - \frac{1}{2\pi}\int_S \Big(2\Im\tau F^{2,0}\wedge F^{0,2}+\sqrt{\Im\tau}\,\calh\,j\wedge F-\frac14\,\calh^2\, j\wedge j\Big)
\ee
Indeed, the action is still invariant under the  action of $\tilde\calq_{\dot\alpha}$ (up to boundary terms) and one can integrate out $\calh$ by setting $\calh=\frac12\sqrt{\Im\tau}\, j^{mn}F_{mn}$ and getting back the formulation  without auxiliary field. For simplicity, we will continue without using the auxiliary field $\calh$.

As observed at the end of section \ref{sec:group}, for constant $\tau$ the above $(0,2)$ TDT reduces  to an ordinary $(2,2)$ topological twist provided by 
twisting  $J$ into $J'\equiv J+ 2\calr$. Indeed, in this subcase the action (\ref{dtwaction}) becomes invariant under two additional twisted supercharges $\calq_\alpha$ given by the complex conjugated of (\ref{dtwsuper}).
See for instance \cite{Dijkgraaf:1997ce} for a detailed discussion on  the $(2,2)$  twisted supersymmetric structure.



\section{Duality walls and chiral defects}
\label{sec:dwc}

As we have already mentioned, in order to appropriately define the four-dimensional action (\ref{dtwaction}) we have to introduce branch cuts, as for instance in \cite{Gaberdiel:1998mv}, along which the theory jumps by an ${\rm SL}(2,\mathbb{Z})$ duality transformation. We call such three-dimensional cuts {\em duality walls}  \cite{Ganor:1996pe,Gaiotto:2008ak,Kapustin:2009av}. In turn, the duality walls can either join together or end on two-dimensional subspaces, around which the theory undergoes an ${\rm SL}(2,\mathbb{Z})$ monodromy. Differently from the duality walls, which do not contain additional degrees of freedom and can be quite freely deformed without affecting the physics, 
 these two dimensional subspaces support two-dimensional chiral theories and for this reason we call them {\em chiral defects}. In this section we discuss the contributions of duality walls and chiral defects to the duality twisted action. Their geometrical structure will be more accurately addressed in section \ref{sec:6d} and a simple explicit example will be provided in section \ref{sec:example}.

\subsection{Duality walls}
\label{sec:dw}

\begin{figure}[h!]
  \centering
    \includegraphics[width=0.28\textwidth]{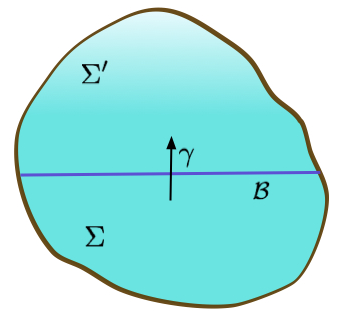}
      \caption{\small A local region of $S$ constituted by two patches $\Sigma$ and $\Sigma'$ divided by a $\gamma$ wall $\calb$. Two of the directions parallel to $\calb$ are suppressed. \label{fig:wall}}
      \end{figure}

Suppose that a local region of $S$ is divided in two parts $\Sigma$ and $\Sigma'$ by a duality wall $\calb$, with orientation such that $\calb=\del\Sigma'=-\del\Sigma$, along which the theory undergoes a duality $\gamma\in {\rm SL}(2,\mathbb{Z})$, see figure \ref{fig:wall}. We then call $\calb$ a $\gamma$ wall. Since $I_{\rm B}$ and $I_{\rm F}$ in (\ref{dtwaction}) are manifestly ${\rm SL}(2,\mathbb{Z})$ invariant and, then, trivially extend across $\calb$, we can focus on $I_{\rm YM}$. It splits in two parts,  $I_{\rm YM}(\Sigma)+I_{\rm YM}(\Sigma')$, which must be somehow glued together.
As discussed in \cite{Ganor:1996pe,Gaiotto:2008ak,Kapustin:2009av}, this requires the addition of a Chern-Simons-like  three-dimensional contribution $I^\gamma_{\rm 3d}$ supported on $\calb$. 
      
      In order to describe $I^\gamma_{\rm 3d}$, let us initially focus on the ${\rm SL}(2,\mathbb{Z})$ generators
\be
T=\left(\begin{array}{cc} 1 & 1\\ 0 &1\end{array}\right)\quad~~~~~~~~~~~~~S=\left(\begin{array}{rr} 0 & -1\\ 1 &0\end{array}\right)
\ee
Suppose first that $\gamma=T$. Then the gauge fields $A$ and $A'$   can be identified and the $T$ wall contributes to the action by the term
\be\label{csT}
I^T_{\rm 3d}=\frac{\ii}{4\pi}\int_\calb A\wedge F
\ee
As a simple check, by computing the equations of motion of $I_{\rm YM}(\Sigma)+I_{\rm YM}(\Sigma')+I^T_{\rm 3d}$ one can easily realize that (\ref{csT}) 
induces the appropriate boundary condition  $F'_{\rm D}|_\calb=(F_{\rm D}+F)|_\calb$ -- cf.~(\ref{dualF}) and (\ref{dualityF}). Analogously, one can easily see that a $T^k$ duality wall, with $k\in\mathbb{Z}$,
supports the contribution  $\frac{ \ii k}{4\pi}\int_\calb A\wedge F$ to the effective action.

Suppose now that the two patches are related by an $S$-duality. In this case, the appropriate boundary term is
\be\label{csS}
I^S_{\rm 3d}=\frac{1}{2\pi\ii}\int_\calb A\wedge F'
\ee
Indeed, it contributes to the equations of motion by the appropriate boundary gluing conditions
\be
F'|_\calb=F_{\rm D}|_\calb\quad~~~~~~~~  F'_{\rm D}|_\calb=-F|_\calb
\ee
The wall associated with $T^{-1}$ and $S^{-1}$ just corresponds to a change of the orientation of $\calb$, that is, to a change of sign on the r.h.s.\ of (\ref{csT}) and (\ref{csS}) respectively. 

Since $T$ and $S$ generate ${\rm SL}(2,\mathbb{Z})$, a more general duality wall can be considered as composite wall which can be `resolved' into  a stack of $T$ and $S$ walls. 
One could also find an explicit expression for the corresponding contribution to the action \cite{Ganor:1996pe} 
\be\label{csgen}
I^{\gamma}_{\rm 3d}=\frac{\ii}{4\pi c}\int_{\calb}(A\wedge F'+ A'\wedge F-d\, A\wedge F-a\, A'\wedge F')
\ee
where the integers $(a,b,c,d)$ define $\gamma$ as in  (\ref{gamma}). Clearly, this formula makes sense only if $c\neq 0$, i.e.\ if $\gamma$ is not of the form $T^k$.
Notice that for $c> 1$ the terms in (\ref{csgen}) have fractional  Chern-Simons-like form, which could be problematic at the quantum level \cite{Kapustin:2009av}.

The above  duality wall terms can be incorporated in a more synthetic formula.  Namely, we can consider $S$ as a space with boundary given  the complete set of duality walls.
Then the overall three-dimensional contribution can be just written as
\be\label{bb}
I_{\rm 3d}=\frac{\ii}{4\pi}\int_{\del S} A\wedge F_{\rm D} 
\ee 
Of course,  each duality wall is the boundary (with opposite orientation) of both regions separated by $\calb$ and then contributes twice. 
In order to recover the previous formulation,   one needs to express each $F_{\rm D}$ in terms of the appropriate combination of elementary gauge fields leaving on the confining patches.
For instance, in the simple  two patches case of figure \ref{fig:wall} considered above $\del S=\del\Sigma\cup\del\Sigma'$, with $\del\Sigma'=-\del\Sigma=\calb$  and (\ref{bb}) is just given by 
\be\label{ba2}
I_{\rm 3d}=\frac{\ii}{4\pi}\int_{\calb}(A'\wedge F'_{\rm D} -A\wedge F_{\rm D}) 
\ee 
For the $T$ wall,  (\ref{csT}) is then recovered by using $A'|_\calb=A|_\calb$, $F'_{\rm D}|_\calb=(F_{\rm D}+F)|_\calb$ in (\ref{ba2}). 
In the case of the $S$ duality wall, (\ref{ba2}) reproduces   (\ref{csS})  by using $F_{\rm D}|_\calb=F'|_\calb$ and $F'_{\rm D}|_\calb=-F|_\calb$ 
 (after an  integration by parts). Finally, in the case of the more general duality wall with $c\neq 0$, in order to recover (\ref{csgen})  from (\ref{ba2}) one needs to use
  $F_{\rm D}|_\calb=\frac{1}{c}(F'-d F)|_\calb$ and $F'_{\rm D}|_\calb=-\frac{1}{c}(F-aF')|_\calb$.\footnote{The  $-\bbone\in{\rm SL}(2,\mathbb{Z})$ duality
 requires some care. Indeed, by making the identification $F'|_\calb=-F|_\calb$ and $F'_{\rm D}|_\calb=-F_{\rm D}|_\calb$, we could have the impression that (\ref{ba2}) identically vanish. However, even though the gauge fields on two patches   are related by a simple sign change, they are still different and  must be treated as independent in the extremization procedure. Since in this case  $F_{\rm D}$ and $F'_{\rm D}$ cannot be eliminated from (\ref{ba2}), this latter cannot be literally applied in this case. Rather, it can be convenient to use $-\bbone =S^2$ and obtain the appropriate wall theory by coalescing two terms of the form (\ref{csS}).}
 
\begin{figure}[h!]
  \centering
    \includegraphics[width=0.28\textwidth]{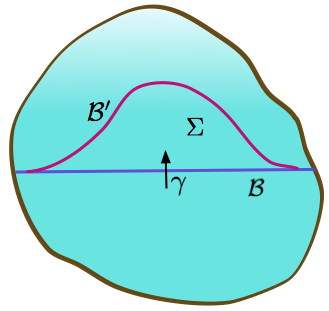}
      \caption{\small A local deformation of a $\gamma$ wall from $\calb$ (blue) to $\calb'$ (violet). It corresponds to performing a $\gamma^{-1}$ duality in the region $\Sigma$ surrounded by $\calb$ and $\calb'$. \label{fig:walldef}}
      \end{figure}

  Let us stress that the choice of duality walls is not unique. Not only they obviously transform after a change of the global duality frame, but they can also be freely deformed, provided the consistency of the overall configuration (to be discussed in the following sections) is preserved. Suppose one wants to deform a $\gamma$ wall from $\calb$ to a nearby $\calb'$, see figure \ref{fig:walldef}. This deformation physically corresponds to performing a $\gamma^{-1}$ duality in the region $\Sigma$ sweeped out in this deformation. (This is actually a way for motivating the presence of the CS-like terms on the duality wall \cite{Gaiotto:2008ak}.) Of course, the new theory obtained after this duality is physically equivalent to the original one.

\subsection{Duality wall junctions}

\begin{figure}[h!]
  \centering
    \includegraphics[width=0.28\textwidth]{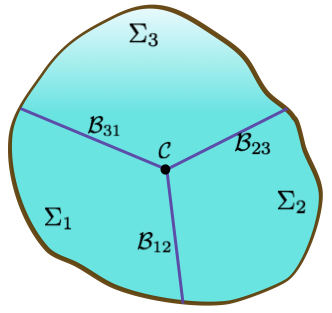}
      \caption{\small A junctions of three duality walls which separates the three patches $\Sigma_1$, $\Sigma_2$ and $\Sigma_3$ and meets at the two-dimensional space $\calc$. The picture shows a 
      two-dimensional slice transversal to $\calc$. \label{fig:junction}}
      \end{figure}

Suppose to have a certain number of patches $\Sigma_\alpha$, which meet on certain 3-manifolds $\calb_{\alpha\beta}=-\calb_{\beta\alpha}$ such that $\del\Sigma_{\alpha}=\sum_{\beta}\calb_{\alpha\beta}$. For concreteness we focus on an ordered set of adjacent  three such patches $\Sigma_1$,  $\Sigma_2$ and $\Sigma_3$, which touch along a certain two-dimensional space $\calc$ -- see figure \ref{fig:junction} -- as more general junctions can be analyzed along the same lines.  The (local) three-dimensional contribution
(\ref{bb}) to the action  then reads
\be\label{3junk}
\frac{\ii}{4\pi}\int_{\calb_{12}}(A^{(1)}\wedge F^{(1)}_{\rm D} -A^{(2)}\wedge F^{(2)}_{\rm D}) +[(1,2)\rightarrow (2,3)]+[(1,2)\rightarrow (3,1)]
\ee
which can be then expressed just in terms of elementary gauge fields $A^{(1)}$, $A^{(2)}$ and $A^{(3)}$ by using the formula (\ref{csgen}) at each wall.
Define $\calc_{\alpha\beta}\equiv \del\calb_{\alpha\beta}=-\calc_{\beta\alpha}$. Then, the consistency conditions $\del (\calb_{12}+\calb_{13})=0$, $\del (\calb_{23}+\calb_{21})=0$ and $\del (\calb_{31}+\calb_{32})=0$ imply that 
\be
\calc_{12}=\calc_{23}=\calc_{31}\equiv \calc
\ee
The theories on $\Sigma_\alpha$ and $\Sigma_\beta$ are related by  a duality  $\gamma_{\alpha\beta}\in {\rm SL}(2,\mathbb{Z})$: $\tau_{\alpha}=\gamma_{\alpha\beta}\cdot\tau_\beta$. Cleary, $\calc$ should not support any non-trivial physical effect.  As a non-trivial consistency check, one should verify that the combination (\ref{3junk}) admits a well defined variational problem,  by checking in particular that possible contributions localized on $\calc$ drop out. This can be indeed verified by 
using the consistency condition $\gamma_{\alpha\beta}\gamma_{\beta\gamma}\gamma_{\gamma\alpha}=\bbone$.

As the simplest example, suppose that all the transformations are just powers of $T$. 
In this case  $A_1=A_2=A_3\equiv A$  and the variation of (\ref{3junk})
produces a two-dimensional term proportional to $(b_{12}+b_{23}+b_{31})\int_\calc \delta A\wedge A$. Indeed, this vanishes by using $\gamma_{\alpha\beta}\gamma_{\beta\gamma}\gamma_{\gamma\alpha}=\bbone$, which boils down to the condition $b_{12}+b_{23}+b_{31}=0$. 
In the same way, though with some more work, one can verify that the same property holds for the cases with two or three non-vanishing $c_{\alpha\beta}$ (only one $c_{\alpha\beta}\neq 0$ is not consistent with $\gamma_{\alpha\beta}\gamma_{\beta\gamma}\gamma_{\gamma\alpha}=\bbone$).\footnote{Walls associated with different dualities generically  intersect transversely along codimension two subspaces. The intersection can be regarded as junction which automatically satisfies the correct monodromy conditions. }

Analogously one can check that monodromy-free junctions do not break gauge invariance. This is easily seen for the most general action case  by using the compact formula (\ref{ba2}) 
for the duality wall contributions.  Take for instance the above three-patches case, whose  wall action is (\ref{3junk}).  Under a gauge transformation $A^{(1)}\rightarrow A^{(1)}+\d\lambda^{(1)}$
\be
\frac{\ii}{4\pi}\int_{\calc_{12}}\lambda^{(1)} F^{(1)}_{\rm D}-\frac{\ii}{4\pi}\int_{\calc_{31}}\lambda^{(1)} F^{(1)}_{\rm D}\equiv 0
\ee
 and, of course,  the same result holds under gauge transformations of $A^{(2)}$ and $A^{(3)}$.

\subsection{Chiral defects}
\label{sec:cdef}

Duality walls can also end on a chiral defect $\calc$ around which the duality twisted theory undergoes a non-trivial ${\rm SL}(2,\mathbb{Z})$ monodromy. 
Clearly, in this case $\calc$ has a physical nature. 
In the F-theory realization, see  section \ref{sec:Ftheory}, the chiral defect $\calc$ represents the intersection of the Euclidean D3-brane and bulk 7-branes. 
 
\begin{figure}[h!]
  \centering
    \includegraphics[width=0.28\textwidth]{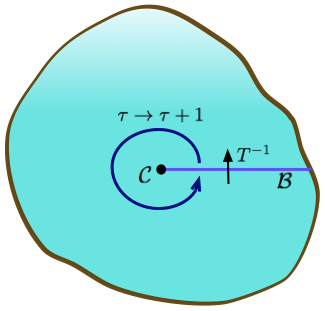}
      \caption{\label{brc}\small Two dimensional slice, transversal to a chiral defect  $\calc$, of a patch $\Sigma$ surrounding $\calc$. As we encircle $\calc$ anticlockwise, $\tau$ undergoes a monodromy $\tau\rightarrow \tau+1$. Hence the wall $\calb$ is associated to the $T^{-1}$ duality which brings $\tau$ back to its original value. }
      \end{figure}

As we will discuss in section \ref{sec:6d}, there are several possible ${\rm SL}(2,\mathbb{Z})$  monodromies, providing  different kinds of chiral defects. However, here we would like to focus on the simplest one, which is associated with a $T$-monodromy around $\calc$ and, in a certain sense,  corresponds to the basic building block for analysing more complicated monodromies.  In such case, the local profile of $\tau$ around $\calc$ is well approximated by $\tau=\frac{1}{2\pi\ii}\log u$, where $u$ is a local complex coordinate transversal to $\calc$ which vanishes on it.  Since, after the loop around $\calc$,  $ \tau$ has shifted by one unit, we need to introduce a $T^{-1}$ wall $\calb$ which makes $\tau$ jump back to its original value, making (\ref{dtwaction}) well defined theory around $\calc$, see figure \ref{brc}.  
According to the discussion of section \ref{sec:dw}, $\calb$ supports  a contribution to the action of the form  
\be\label{cschiral}
I_{\rm 3d}=\frac{1}{4\pi\ii}\int_\calb A\wedge F
\ee 

Now,  this action is clearly not invariant and  under the gauge transformation $A\rightarrow A+\d\lambda$ since it  produces an anomalous term localized along $\calc$:
\be\label{3danomaly}
\Delta I_{\rm 3d}=\frac{1}{4\pi\ii}\int_\calc \lambda F
\ee
Such a term has precisely the right form to be cancelled by the anomaly produced by a two-dimensional chiral fermion  localized on $\calc$. In order to clearly see this, it is convenient to  adopt
the dual bosonized description of the chiral fermion. This uses a self-dual scalar $\beta$, constrained by $*_\calc\d\beta=\ii\d\beta$, which does not admit a standard Lagrangian description. Rather, one can use the off-shell action
 \be\label{2dB}
\begin{aligned}
I_{\rm 2d}=&\frac{1}{8\pi}\int_\calc *(\d\beta-A)\wedge(\d\beta-A)+\frac{\ii}{4\pi}\int_\calc\beta F
\\=&\frac\ii{4\pi}\int_\calc \Big(\del\beta\wedge\delbar\beta-2\del\beta\wedge A^{0,1}+A^{1,0}\wedge A^{0,1}\Big)
\end{aligned}
\ee
for the {\em unconstrained} scalar and then picks-up the contribution  from the chiral part directly at the level of the partition function, as described in detail in \cite{Witten:1996hc}. 
Even if subtle in this respect, the action (\ref{2dB}) produces the anomaly of the chiral theory already at the classical level. Indeed, under a gauge transformation the scalar shifts by $\beta\rightarrow \beta+\lambda$ and one can immediately check that (\ref{2dB}) produces an  anomalous term which perfectly cancels (\ref{3danomaly}). Clearly, if needed, one can always go to the fermionic formulation.\footnote{The bosonic formulation requires that, in absence of additional insertions, $F$ trivializes along $\calc$. The interpretation is that  the partition function vanishes for  non-trivial  $F|_\calc$. This is clear by using the fermionic formulation, in which the integer $n=\frac1{2\pi}\int_\calc F$ counts the number of fermionic zero-modes on $\calc$, which makes the partition function vanish in absence of proper insertions.}

Of course, our characterization of the above chiral defect in terms of a $T$-monodromy depends on the choice of duality frame.\footnote{Furthermore, notice that  the monodromy in general depends on the base point, that is, the starting point of the path along which one computes the monodromy. In particular, if one moves the base point through a duality wall, the monodromy changes to its conjugated by the duality associated with the wall.} Indeed, by applying a duality $M\in{\rm SL}(2,\mathbb{Z})$,  the monodromy associated with chiral defect becomes
\be\label{conjmon}
\gamma_{[p,q]}\equiv M\left(\begin{array}{cc} 1 & 1 \\ 0 & 1 \end{array}\right)M^{-1}
=\left(\begin{array}{cc} 1-pq & p^2 \\ -q^2 & 1+pq \end{array}\right)
\ee
with
\be
 M=\left(\begin{array}{cc} p & r \\ q & s \end{array}\right)\in{\rm SL}(2,\mathbb{Z})
\ee
As suggested by the notation, this monodromy can be completely characterized in terms of a pair of integers $[p,q]$, such that $ps-qr=1$ for some other two integers $r,s$.\footnote{The ambiguity in the choice of $r,s$ is due to the possibility to substitute $M$ with $MT^k$ for an arbitrary integer $k$.} In particular, this implies that $p$ and $q$ must be relatively prime. 
In this way we obtain what we call a $[p,q]$-chiral theory. The theory described by (\ref{2dB}) then corresponds to a $[1,0]$ chiral theory. 

\begin{figure}[h!]
  \centering
    \includegraphics[width=0.60\textwidth]{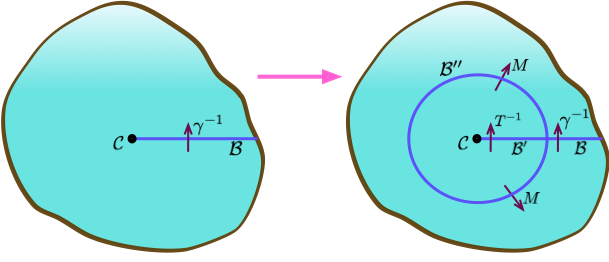}
      \caption{\label{res}\small A $T$-conjugated $\gamma$-monodromy, with $\gamma=M T M^{-1}$, can be locally described as a  $T$-monodromy by appropriately choosing the duality walls. The $\gamma^{-1}$ wall $\calb$ on the left 
      is substituted by a junction of the three $\gamma^{-1}$, $T^{-1}$ and $M$  walls ($\calb$, $\calb'$ and $\calb''$ respectively) on the right.}
\end{figure}

By construction the  $[p,q]$-chiral theory is anomalous under the bulk gauge transformations, with an anomaly which cancels the anomaly produced by the three-dimensional theory supported by the duality wall $\calb$.
However, its direct description is not obvious. Then, it could be useful to adjust the duality walls to `resolve' a $\gamma_{[p,q]}$-monodromy into a $T$-monodromy as in figure \ref{res}. 
This description of the chiral defect  could be not  possible globally. In fact, such complications are expected to be generically present.

\subsection{Supersymmetry revisited}
\label{sec:susy}

Let us summarize what we have discussed so far. In order to appropriately define the four-dimensional action (\ref{dtwaction}), one must introduce a network of duality walls. These support the three-dimensional contribution  (\ref{bb}) to the action. The  walls can end on chiral defects around which the theory experiences ${\rm SL}(2,\mathbb{Z})$ monodromies and which give additional two-dimensional contributions to the action. In particular, in the basic case of $T$-monodromy, i.e.\ of $[1,0]$ chiral defect,  one has to add (\ref{2dB}) to the action. The presence of such a structure was basically already observed in \cite{Ganor:1996pe}.   

We can now revisit the duality twisted supersymmetry of the overall system. Let us first focus on the local two-patches system depicted in figure \ref{fig:wall}.
It is described  by the action 
\be\label{2pg}
I_{\rm 4d}(\Sigma)+I_{\rm 4d}(\Sigma')+I_{\rm 3d}(\calb)
\ee
where the different terms are given in (\ref{dtwaction}) and (\ref{ba2}).  Since $\del\Sigma'=-\del\Sigma=\calb$ the variation of $I_{\rm 4d}(\Sigma)+I_{\rm 4d}(\Sigma')$ under the supersymmetry transformation $\delta=\zeta_{\dot\alpha}\tilde\calq^{\dot\alpha}$ is given by
\be\label{bulkvar}
\delta I_{\rm 4d}(\Sigma)+\delta I_{\rm 4d}(\Sigma')=\frac{1}{2\pi\ii} \int_\calb\zeta_{\dot\alpha}\Big(  \frac{\tau'}{\sqrt{\Im\tau'}}\,\tilde\psi'^{\dot\alpha}\wedge F'- \frac{\tau}{\sqrt{\Im\tau}}\,\tilde\psi^{\dot\alpha}\wedge F \Big)
\ee
We should check whether such a term is compensated by $\delta I_{\rm 3d}(\calb)$. 

Consider first the case of a $T$-duality wall: $\tau'=\tau+1$. Then $I_{\rm 3d}$ is given by (\ref{csT})
whose supersymmetry variation under $\delta=\zeta_{\dot\alpha}\tilde Q^{\dot\alpha}$ is
\be
\delta I_{\rm 3d}=\frac{\ii}{2\pi}\int_{\calb}\frac{1}{\sqrt{\Im\tau}}\,\zeta_{\dot\alpha}\tilde\psi^{\dot\alpha} \wedge F
\ee
This exactly cancels the contribution from (\ref{bulkvar}). 

On the other hand, consider an $S$ wall, supporting (\ref{csS}). Taking into account that $\tau'=-1/\tau$, 
$\delta A=(\Im\tau)^{-\frac12}\zeta_{\dot\alpha}\tilde\psi^{\dot\alpha}=-\tau'(\Im\tau')^{-\frac12}
\zeta_{\dot\alpha}\tilde\psi'^{\dot\alpha}$ and 
$\delta A'= (\Im\tau')^{-\frac12}\zeta_{\dot\alpha}\tilde\psi'^{\dot\alpha}=\tau(\Im\tau)^{-\frac12}\zeta^{\dot\alpha}\tilde\psi_{\dot\alpha}$ one finds that $\delta I^{S}_{\rm 3d}$ exactly cancels (\ref{bulkvar}). One can actually repeat the same calculations for more general duality walls supporting the term (\ref{csgen}).

Let us now examine the contributions from possible chiral defects $\calc\equiv \del\calb$, focusing on the $[1,0]$ chiral defects,  around which the theory undergoes a $T$-monodromy. As discussed above, more general $[p,q]$ chiral defects can be locally reduced to this case by adjusting the duality wall network.  The three-dimensional term (\ref{cschiral})  produces a contribution localized on $\calc$. Indeed, by varying (\ref{cschiral}) under $\delta=\zeta_{\dot\alpha}\tilde \calq^{\dot\alpha}$, integration by parts produces a two-dimensional anomalous contribution
\be\label{branchcontr}
 \delta I_{\rm 3d}=\frac{1}{4\pi\ii}\int_\calc  \frac{1}{\sqrt{\Im\tau}}\,\zeta_{\dot\alpha}\tilde\psi^{\dot\alpha}\wedge A
\ee
It is now easy to check that this is exactly cancelled  by the twisted supersymmetry variation  of the two-dimensional action (\ref{2dB}), where the chiral boson is completely inert under the twisted supersymmetry.

Hence, we see that four-, three- and two-dimensional terms in the duality twisted action  conspire to obtain a theory which preserves the twisted supercharges (\ref{dtwsuper}).

 A final comment on the role of the two-dimensional chiral boson. In the above manipulations we have use the embedding of the chiral boson in the non-chiral theory (\ref{2dB}). On the other hand, the chiral theory coupled to an exterior gauge field $A$ can be defined at the level of partition  function $Z_{\rm 2d}(A)$ \cite{Witten:1996hc}  obtained  by `throwing away' the non-chiral contribution, and  so one may wonder if the above manipulations remain true just for the chiral theory.  However, it is sufficient to recall that the partition function $Z_{\rm 2d}(A)$ can be considered as a holomorphic section of a line bunde on the space of gauge fields, on which one declares that $A^{0,1}$ are holomorphic while $A^{1,0}$ are anti-holomorphic `coordinates' \cite{Witten:1996hc}. Then, the holomorphy of the section corresponds to  the differential equation 
\be\label{2dhol}
\frac{D}{D A^{1,0}}\,Z_{\rm 2d}=0
\ee
  with 
  \be
  \frac{D}{D A^{1,0}}=\frac{\delta}{\delta A^{1,0}}+\frac{\ii}{4\pi}A^{0,1}
  \ee
  where we define the functional derivative  by $\delta Z_{\rm 2d}=\int_\calc\delta A\wedge \frac{\delta Z_{\rm 2d}}{\delta A}$. Of course (\ref{2dhol})
  is satisfied by $e^{-I_{\rm 2d}}$ too, with $I_{\rm 2d}$ as in (\ref{2dB}). The key point is that  (\ref{2dhol}) is everything we need for the cancellation of  (\ref{branchcontr}) (as well as of the anomaly (\ref{3danomaly})), which is then still valid by using the two-dimensional chiral partition function instead of $e^{-I_{\rm 2d}}$.

\section{Global aspects}
\label{sec:6d}

Our TDT requires a holomorphic coupling $\tau$ over the space $S$, which can undergo non-trivial ${\rm SL}(2,\mathbb{Z})$ duality as one moves about. As in F-theory, the global structure of such a $\tau$ can be conveniently described by using an auxiliary elliptic fibration.  In this section we revisit a few aspects of  this strategy, which are well-known in the F-theory context, see for instance \cite{Denef:2008wq,Weigand:2010wm,Taylor:2011wt} for reviews, adapting them to our problem.

\subsection{General considerations}
 
 Let us  consider  a six-dimensional complex manifold  $D$ which is elliptically fibered over $S$. If $\pi:D\rightarrow S$ denotes the projection map, $\tau(z)$ corresponds to the complex structure of the elliptic fiber $\pi^{-1}(z)$.  Physically, the duality twisted theory on $S$ corresponds to the four-dimensional  effective theory obtained by compactifying the $(2,0)$ six-dimensional theory on $D$ in the fiber zero-size limit.

The elliptic fibration is characterized by the holomorphic line bundle $\call_\cald$ which has played an important role in our construction. 
Indeed, one can use $\call_\cald$ for explicitly defining the elliptic fibration  by a Weierstrass equation
\be\label{weiss}
y^2=x^3+f(z)x+g(z)
\ee
where $f$ and $g$ are sections of $\call^{4}_\cald$ and $\call_\cald^6$ respectively.\footnote{Let us emphasize an important difference with respect to the more familiar F-theory construction: the line bundle $\call_\cald$ and the anti-canonical bundle $K_S^{-1}$ are generically independent.  This is because gravity is treated as an external non-dynamical field over which the non-trivial $\tau$ profile does not backreact. Instead, in a complete F-theory background, the backreaction of $\tau$ on the metric forces the anti-canonical bundle to be isomorphic to $\call_\cald$.}

In this language, the chiral defects characterizing the TDT are localized at the (holomorphic) curves $\calc_i$ on which  the elliptic fiber degenerates. Let us recall that the discriminant of the Weierstrass model is defined as
\be
\Delta=4f^3+27 g^2
\ee
and is then a section of $\call_\cald^{12}$. The chiral defects are localized on the divisor
\be\label{Cdiscr}
\calc=\sum_in_i\calc_i=\{\Delta=0\}
\ee
where $\calc_i$ denotes the $i$-th irreducible component curve and $n_i$ its multiplicity. 

The holomorphic $\tau(z)$ can be implicitly identified, up to modular ${\rm SL}(2,\mathbb{Z})$ transformations, in terms of $f(z)$ and $g(z)$, through the modular invariant $j(\tau)$-function:
\be\label{jfunct}
j(\tau)= \frac{4(12 f)^3}{\Delta}
\ee
which has a zero at $\tau=e^{\frac{\pi\ii}{3}}$, diverges as $\tau\rightarrow \ii\infty$ and is normalized so that $j(\ii)=(12)^3$.
More precisely, for $\Im\tau\gg 1$ we have $j(\tau)=e^{-2\pi\ii\tau}+744+\calo(e^{2\pi\ii\tau})$.

On can actually have various possible fiber degenerations, which were classified by Kodaira \cite{kodairaII,kodairaIII} in terms of the vanishing order  of $\Delta$, $f$ and $g$. In particular, each type of degeneration is characterized by a certain  singularity of the total space $D$ and a certain ${\rm SL}(2,\mathbb{Z})$ monodromy, which is defined up to an ${\rm SL}(2,\mathbb{Z})$ conjugation.  Kodaira's classification of minimal degenerations is  summarized  in table \ref{monodromies}. The last column  of the table gives the  Lie algebra whose Cartan matrix describes the intersection numbers of the $\mathbb{P}^1$'s which must be blown up to resolve the singularity. In addition there are non-minimal degenerations, with ${\rm ord}_{\calc}(\Delta)\geq 12$, ${\rm ord}_{\calc}(f)\geq 4$ and ${\rm ord}_{\calc}(g)\geq 6$, which we discard, as is usually done  in F-theory on physical grounds -- see also \cite{McOrist:2010jw} for a discussion from a viewpoint similar to the one adopted in this paper. 

\begin{table}
\centering
{\footnotesize\begin{tabular}{ | c | c | c | c | c | c |}
  \hline 
   &
     ${\rm ord}_{\calc}(\Delta)$ &${\rm ord}_{\calc}(f)$ & ${\rm ord}_{\calc}(g)$ & {\rm monodromy} & {\rm singularity} \\
     \hline  
     \hline
      I$_0$ & 0 & $\geq 0$ & $\geq 0$ & $\left(\begin{array}{cc} 1 & 0 \\ 0 & 1 \end{array}\right)$  & none  \\
       \hline
      I$_n$, $n\geq 1$ & $n$ & 0 & 0 & $\left(\begin{array}{cc} 1 & n \\ 0 & 1 \end{array}\right)$  & $A_{n-1}$  \\
        \hline
      II & $\geq 2$ & 1 & 1 & $\left(\begin{array}{rr} 1 & 1 \\ -1 & 0 \end{array}\right)$  & none  \\
      \hline
      III & 3 &1 & $\geq 2$ &  $\left(\begin{array}{rr} 0 & 1 \\ -1 & 0 \end{array}\right)$  & $A_1$  \\
        \hline
      IV& 4 & $\geq 2$ & $2$ &  $\left(\begin{array}{rr} 0 & 1 \\ -1 & -1 \end{array}\right)$  & $A_2$  \\
        \hline
      I$_0^*$&  6 & $\geq 2$ & $\geq 3$ & $\left(\begin{array}{rr} -1 & 4 \\ 0 & -1 \end{array}\right)$  & $D_4$  \\
       \hline
       I$^*_n$, $n\geq 1$ &$6+n$ & $2$ & $3$ &  $\left(\begin{array}{rr} -1 & 4 \\ 0 & -1 \end{array}\right)$  & $D_{4+n}$  \\
       \hline
       IV$^*$ &$8$ &  $\geq 3$ & $4$ &  $\left(\begin{array}{rr} -1 & -1 \\ 1 & 0 \end{array}\right)$  & $E_6$  \\
       \hline
       III$^*$ & $9$ & $3$ & $\geq 5$ &  $\left(\begin{array}{rr} 0 & -1 \\ 1 & 0 \end{array}\right)$  & $E_7$  \\
        \hline
       II$^*$ &$10$ &  $\geq 4$ & $5$ &  $\left(\begin{array}{rr} 0 & -1 \\ 1 & 1 \end{array}\right)$  & $E_8$  \\
    \hline  
\end{tabular}
\caption{Kodaira's classification of fiber degenerations.}\label{monodromies}}
\end{table}

We see that the $[p,q]$ chiral defects considered  in the previous section correspond to type I$_1$ degenerations, along which the elliptic fiber degenerates to a rational curve with a double point. We know that, in the appropriate ${\rm SL}(2,\mathbb{Z})$ frame, they support
 a two-dimensional chiral theory of the kind discussed in subsection \ref{sec:cdef}. This must correspond to a contribution of the six-dimensional chiral two-form localized at the degeneration locus \cite{Donagi:2010pd,Clingher:2012rg}. Roughly, one can  locally construct an anti-self-dual two-form $\omega_{\it 2}$ localized around the I$_1$ singularity as in section 3.9 of \cite{Denef:2008wq}. One can then expand the self-dual field-strength $T_{\it 3}$ of the six-dimensional theory as  $T_{\it 3}=\d\beta\wedge\omega_{\it 2}$, hence obtaining our chiral boson $\beta$ localised on $\calc$. By fermionization, one then  identifies $\frac1{2\pi}\d\beta$ with the two-dimensional chiral current $j_\calc=\bar\chi\chi$, where $\chi$ is a chiral fermion. 
 
Along more general  degenerations the fiber splits into a tree of $\mathbb{P}^1$'s, call them $C_k$, with $k=0,\ldots,n$, where $n$ denotes the rank of Lie algebra provided by last column of table \ref{monodromies}. The $C_i$'s intersections give the associated extended Cartan matrix. Then, one can naturally integrate $T_{\it 3}$ over these two-spheres and combinations thereof obtaining a set of two-dimensional chiral currents localized along the degeneration curve $\calc$, transforming under the group associated with the last column of table \ref{monodromies}. A similar argument can be applied for enhanced degenerations at the intersections of the degeneration curves. These aspects will not be further developed in this paper, as for simplicity we will be mostly concerned with I$_1$ degenerations.

Finally, as in F-theory, there is a particular regime in which the intricacies of the general duality twisted configuration can be made more tractable. Namely, following \cite{Sen:1996vd,Sen:1997gv}, one can consider  the so-called Sen limit, in which the coupling $\tau$ can be made almost constant and arbitrary small. The way to adapt this limit to the  topologically twisted theory considered in this paper follows  quite straightforwardly from \cite{Sen:1997gv}. Hence, instead of presenting a general discussion about this limit, we will work it out  in the concrete example presented in the following subsection.


\subsection{An example}
\label{sec:example}

As a concrete example, let us now put the theory on a Hirzebruch surface, $S\equiv F_n$, which is  a $\mathbb{P}^1$ fibration over $\mathbb{P}^1$.\footnote{See for instance \cite{Morrison:1996na} for a short account on Hirzebruch surfaces.}
We denote the fiber by $\mathbb{P}^1_f$ and the base by $\mathbb{P}^1_b$. One can describe $F_n$ by using four coordinates $(s,t,u,v)$ in $\mathbb{C}^4-Z$, with $Z=\{u=v=0\}\cup\{s=t=0\}$, modded out by the identification
$(s,t,u,v)\simeq (\lambda_1s,\lambda_1t,\lambda_2u,\lambda_2\lambda^{-n}_1v)$, where $(\lambda_1,\lambda_2)\in\mathbb{C}^{*2}$.  $(u,v)$ are projective coordinates of $\mathbb{P}^1_f$ while $(s,t)$ are projective coordinates of $\mathbb{P}^1_b$. These coordinates allow the identification of a natural set of curves (divisors) $\calc_s$, $\calc_t$, $\calc_u$ and $\calc_v$ defined by the vanishing of the corresponding coordinate. In homology we have $\calc_t\simeq \calc_s$ and $\calc_u\simeq \calc_v+n\calc_s$ and the non-vanishing intersection numbers between these curves are $\calc_s\cdot\calc_v=1$ and $\calc_v\cdot\calc_v=-n$.  $\calc_s$ and $\calc_t$ can be identified with the (homologous) copies of the fiber $\mathbb{P}^1_f$ attached at the base points $s=0$ and $t=0$ respectively. On the other hand, the non-trivial self-intersection of $\calc_v$ is a manifestation of the non-triviality of the fibration for $n\geq 1$ (while $F_0=\mathbb{P}^1_f\times \mathbb{P}^1_b$).

In order to describe a duality twisted theory on $F_n$, we have to choose the line bundle $\call_\cald$. The most general line bundle on $F_n$ can be constructed as the product of line bundles $\calo_{F_n}(\calc_s)$ and $\calo_{F_n}(\calc_v)$, associated with $\calc_s$ and $\calc_v$. Since $\calc_s\cdot\calc_s=0$, the simplest choice is 
\be
\call_\cald=\calo_{F_n}(\calc_s)
\ee
 and then $f$, $g$ and $\Delta$ are sections of
 $\calo_{F_n}(4\calc_s)$, $ \calo_{F_n}(6\calc_s)$ and $\calo(12\calc_s)$ respectively. A section of  $\calo(k\calc_s)$ can be described as a homogeneous polynomial of degree $k$ in $(s,t)$ and then $f$, $g$ and $\Delta$ are homogeneous polynomial of degree 4,6 and 12 respectively in $\mathbb{P}^1_b$.  Introducing the inhomogeneous coordinate $z=s/t$ along the base, we can write
\be\label{deltaroots}
\Delta=\prod^{12}_{I=1}(z-z_I)
\ee
where $z_I$ are the twelve roots of $\Delta=0$. 

Hence $\calc$ as defined in (\ref{Cdiscr}) can be identified with 12 copies of $\mathbb{P}^1_{f}$ sitting at different points $z_I$ of the base.  More formally, $\calc=\sum_I\calc_I$, with  $\calc_{I}\equiv \pi^*(z_I)$, where $\pi:F_n\rightarrow \mathbb{P}^1_b$ is the projection map of the fibration. 
  Generically $f$ and $g$ are non-vanishing at the $z_I$'s and then, from table \ref{monodromies}, there are twelve I$_1$ chiral defects which wrap the fiber $\mathbb{P}^1_f$  and stay at the points $z_I$ in the base.   These points can be connected by lines along the base, which meet at certain points, forming a network. By attaching the fiber $\mathbb{P}^1_f$ along each line, this network uplifts to a  network of duality walls in $F_n$.
   
 In order to understand  the structure of the duality walls, one needs to understand more precisely the  monodromies around 
 each $z_I$. This problem simplifies considerably if one considers the Sen's limit of the theory \cite{Sen:1996vd,Sen:1997gv}.
 This is achieved by tuning the profile of $\tau$ so that  $f$ and $g$ get the form 
 \be\label{senansatz}
f=-3h^2+\varepsilon \eta\quad~~~~~ g=-2h^3+\varepsilon h\eta-\varepsilon^2\chi
\ee
where $\varepsilon$ is a constant and $h$, $\eta$ and $\chi$ are sections of $\call_\cald^2$, $\call_\cald^4$ and $\call_\cald^6$ that is, they are polynomial in $z$ of degree 2, 4 and 6 respectively. Then
\be\label{sendelta}
\Delta=\varepsilon^2(-9\eta^2h^2-108h^3\chi+4\varepsilon\eta^3+54 \varepsilon h\eta\chi+27\varepsilon^2\chi^2)
\ee
If one takes $\varepsilon\ll 1$ and we are away from the  $h=0$ locus, then  $\Delta\simeq \varepsilon^2$ and then $j\sim \varepsilon^{-2}\gg 1$. Hence, in this region we can approximate $\tau\simeq -\frac{1}{2\pi\ii}\log \varepsilon^2$, so that $\Im\tau\gg 1$ and the theory is almost everywhere weakly coupled.
In the  $\varepsilon\ll 1$ limit, the twelve roots $z_I$ appearing in (\ref{deltaroots})  split in two groups. 

  The first group is obtained by assuming finite $\eta,h,\chi$. Hence one can identify eight approximate roots $z_{A,i}$, $i=1,\ldots,8$,  as the solutions of  $\eta^2+12h\chi=0$, which is an equation of degree eight in $z$. Correspondingly, there are eight curves  $\calc_{A,i}$, obtained by attaching the fiber $\mathbb{P}^1_f$ at the points $z_{A,i}$ in the base. In an appropriate duality frame, each of these curves  supports a $[1,0]$ chiral defect. 
 
 The second group is obtained by considering $h$ very small too (of order $\sqrt{\varepsilon}$), while keeping $\eta$ and $\chi$ finite. Hence, one can find four approximate roots by solving  the degree four equation  $9h^2-4\varepsilon\eta=0$. Since we are assuming that $\varepsilon\ll1$, these four roots are approximately identified as follows. First denote by $e_1,e_2$ the two roots of $h=0$, so that $h=(z-e_1)(z-e_2)$. Hence, the four roots of $9h^2-4\varepsilon\eta=0$ split in two pairs $(z_{B,a},z_{C,a})$, $a=1,2$, 
 localized at a `distance' of order $\sqrt{\varepsilon}$ from  $e_a$. Notice that this distance if exponentially suppressed if expressed terms of the bulk YM coupling constant $\Im\tau\gg 1$.
 Following  \cite{Sen:1997gv}, one can
 argue that, in the frame in which $\calc_{A,i}$ support $[1,0]$ chiral defects, the pair of monodromies  with common base point around the points $z_{B,a}$ and $z_{C,a}$ have the following general form:
 \be
 \gamma_{B}=T^{k-1}S^{-1}T^{-k-1}
 \quad,\quad  \gamma_{C}=T^{k+1}S^{-1}T^{-k-3}
 \ee       
where $k$ is an arbitrary integer, so that $\gamma_{B}\gamma_{C}=-T^{-4}$. Different choices of $k$ are related by a conjugation with an appropriate power of $T$ which, on the other hand, does not modify the $T$-monodromy around the $[1,0]$ defects $\calc_{A,i}$.

\begin{figure}[h!]
  \centering
    \includegraphics[width=0.9\textwidth]{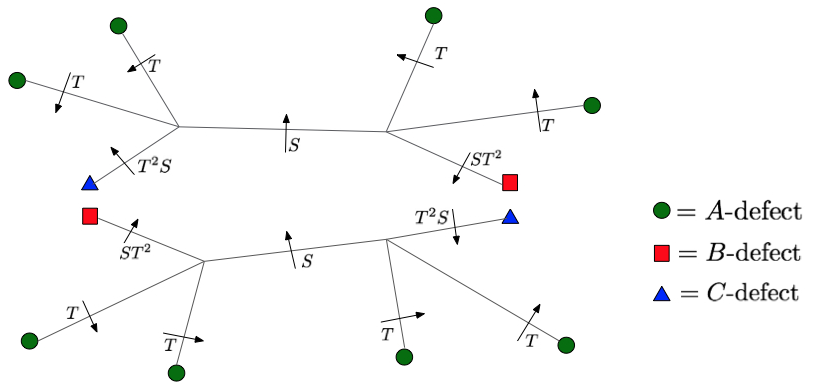}
      \caption{\label{fig:net}\small Example of network of duality walls and chiral defects  on $F_n$ corresponding to the choice $\call_\cald=\calo_{F_n}(D_s)$ at weak coupling. Only the two-dimensional projection to the $\mathbb{P}^1_b$ base is shown (by using the inhomogeneous coordinate $z=s/t$) and the actual defects and walls are obtained by attaching the   $\mathbb{P}^1_f$ fiber at each point.   The (anticlockwise) monodromy around each defects can be recovered by compensating the duality jumps generated by the walls: $\gamma_A=T$, $\gamma_B=T^{-2}S^{-1}$ and $\gamma_C=S^{-1}T^{-2}$.
    }
\end{figure}

The actual value of $k$ is related to the choice of  duality walls, which is not unique. For instance, we can take 
\be
h=z^2-1\quad~~~~~~~~\eta=z^4+8
\ee
and $\chi$ a moderately small constant. Then, the location of the chiral defects with monodromies $\gamma_B$ and $\gamma_C$ corresponding to $k=-1$  as well as  an explicit choice of duality walls, is schematically represented in figure \ref{fig:net}. 
With this choice a $B$ defect corresponds to a $[-1,1]$-defect, while a $C$ defect corresponds to a $[1,1]$-defect. 
As described  in  more generality in section \ref{sec:cdef}, see figure \ref{res}, the $ST^2$ and $T^2S$ walls which terminate on $B$ and $C$ defects can be further resolved into $T^{-1}$ walls surrounded by an $M_B$ and $M_C$ walls, with
\be
M_B=\left(\begin{array}{rr} 1 & 1 \\ -1 & 0 \end{array}\right)\quad~~~~~~~~~~M_C=\left(\begin{array}{rr} 1 & -1 \\ 1 & 0 \end{array}\right)
\ee

\begin{figure}[h!]
  \centering
    \includegraphics[width=0.9\textwidth]{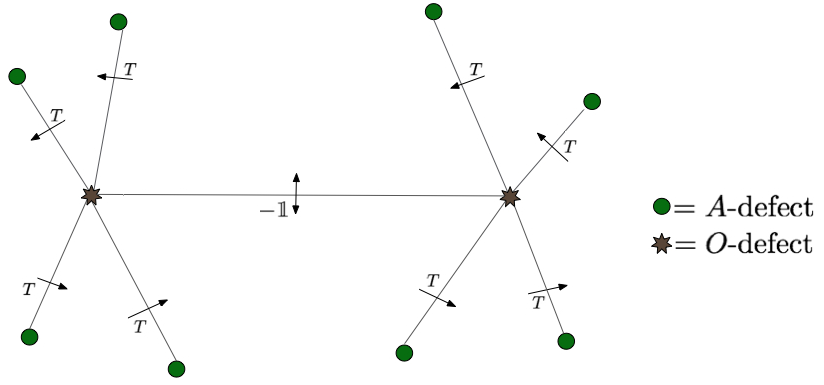}
      \caption{\label{fig:net2}\small Strict weak coupling limit of the system depicted in figure \ref{fig:net}. Each pair of $B$ and $C$ defects has collapsed to an $O$-defect and the two $S$-walls have been merged in a single $-\bbone$ wall, so that each of the two $O$-defects acts as a sink for the $T$-walls sourced by four $A$-defects.}
\end{figure}

We can also explicitly see what happens in the strict weak coupling limit, in which $\varepsilon\rightarrow 0$.\footnote{A refined description of this limit has been recently proposed in \cite{Clingher:2012rg}.} The pairs of chiral defects of type $B$ and $C$ collapse into a pair of degenerate singular loci, located at the two roots of $e_1$ and $e_2$, which we call O-defects, around which the total monodromy is $\gamma_O=-T^{-4}$.  Notice that this monodromy requires that $A\rightarrow -A$ as one encircles  $\calc_{O}$, which means the field strength must satisfy the condition $F|_{\calc_{O}}=0$. The O-defects are connected by a $-\bbone$ wall and each of them acts as a sink for four $T$-walls which departure from $[1,0]$-chiral defects located at the points $z_{A,i}$, see figure \ref{fig:net2}.

Finally, one can also go to a double cover description, which is analogous to the type IIB double cover orientifold description. This can be done by adding one coordinate $\xi$ which transforms as a section of $\call_\cald$ and  define the double  cover $\tilde S$  by the equation
\be\label{basecond}
\xi^2-h(z)=0
\ee
Hence, in this case $\tilde S$ can be regarded as a $\mathbb{P}^1_f$ fibration over a one-dimensional base which is the  double cover of the base $\mathbb{P}^1_b$ of $S$, with the two O-defects as branch points and the $-\bbone$ wall as  branch cut. Every point $z$ with $h(z)\neq 0$ in the base $\mathbb{P}^1_b$ of $S$ uplifts to two points $(z,\xi_\pm)$, with $\xi_\pm=\pm\sqrt{h(z)}$ in the base of $\tilde S$. Hence in $\tilde S$ there are sixteen $A$-defects which are symmetric with respect to the orientifold involution $\xi\rightarrow -\xi$, whose fixed points correspond to the two O-defects. 

The choice $\call_\cald=\calo_{F_n}(\calc_s)$ discussed so far is particularly simple, in particular because the corresponding chiral defects do not intersect.
Of course, one could analyse more general choices for $\call_\cald$, with systems of intersecting chiral defects and more intricate networks of duality walls, but we are not going to address such issues  in this paper.

\section{Embedding in F-theory}
\label{sec:Ftheory}

So far we have discussed the topological duality twist of the U(1) $N=4$ SYM without referring too much to our original motivation 
for discussing such a system, namely the study of supersymmetric  Euclidean D3-branes (E3-branes, for short) instantons in F-theory compactifications to four-dimensions, see  \cite{Denef:2008wq,Weigand:2010wm} for reviews.

Such brane instantons are often studied by using the dual M-theory viewpoint,  in which they correspond to Euclidean M5-branes \cite{Witten:1996bn}. The M-theory viewpoint is particularly useful for addressing global topological issues. On the other hand, the complete supersymmetric M5-brane effective action is purely understood, in particular in presence of non-trivial supergravity backgrounds. In such case the only available candidate is the action in \cite{Bandos:1997ui}, which does not however appear particularly manageable. In this respect, using directly the E3-brane in IIB is a valuable alternative, see for instance \cite{Ganor:1996pe,Cvetic:2009ah,Cvetic:2010rq,Cvetic:2010ky,Cvetic:2011gp,Bianchi:2011qh,Bianchi:2012pn,Bianchi:2012kt} for previous works emphasising the relevance of the IIB viewpoint. Furthermore, in this way one can have a more direct link to the results obtained  in weakly coupled regime, see \cite{Blumenhagen:2009qh,Bianchi:2009ij,Bianchi:2012ud} for reviews. 

 A short summary of the structure of F-theory compactifications from the IIB perspective, in the notation used in the present paper, can be found in sections 2 and 3 of \cite{Bianchi:2012kt}.\footnote{To completely match the conventions of \cite{Bianchi:2012kt} one has to make the sign flip $B_{\it 2}\rightarrow -B_{\it 2}$. Furthermore, the symbols $\call_\cald$, $q_\cald$ and $\cala$ (and relatives) used in the present paper correspond to $\call_Q$, $q_Q$ and $Q$ in \cite{Bianchi:2012pn,Bianchi:2012kt}.}
In such backgrounds, the type IIB space-time has the form $\mathbb{R}^4\times X$, where $X$ is a six-dimensional K\"ahler space. 
The four-dimensional space $S$ represents a K\"ahler submanifold of $X$, the four-dimensional K\"ahler $j$ is given by the pull-back of the bulk K\"ahler form $J$ and the SYM point-dependent coupling $\tau$ just corresponds to the restriction of the bulk axio-dilaton $\tau\equiv C_{\it 0}+\ii e^{-\phi}$ to $S$. Furthermore, the line bundle $\call_\cald$ is the restriction to $S$ of a bulk line bundle, which we denote in the same way. Bulk supersymmetry then imposes that such line bundle is isomorphic to the anticanonical bundle $K_X^{-1}$ of $X$. Hence, on the E3-brane, $\call_\cald\simeq K^{-1}_X|_S\simeq K^{-1}_S\otimes N_{S/X}$.

The four-dimensional fermionic spectrum has been derived in detail in \cite{Bianchi:2011qh,Bianchi:2012kt} starting from the Green-Schwarz formulation of the fermionic D3-brane effective action \cite{Marolf:2003ye,Marolf:2003vf,Martucci:2005rb} and matches the twisted fermions introduced in section \ref{sec:TDT}. This confirms that the arguments of \cite{Bershadsky:1995qy} extend to the F-theory case by using the TDT. On the other hand the bosons $\varphi^{\alpha\dot\beta}$ describe the fluctuations of the E3-brane along the external $\mathbb{R}^4$, while $\sigma$ and $\tilde\sigma$ describe the fluctuations in $X$. More precisely, $\sigma=\frac12 e^{-\frac{\phi}{2}}(\iota_{{\bf v}}\Omega)\varphi^{{\bf v}}$ and $\tilde\sigma=\frac12 e^{-\frac{\phi}{2}}(\iota_{\bar{\bf v}}\bar\Omega)\varphi^{\bar{\bf v}}$, where $e^{-\frac{\phi}{2}}\Omega$ is the globally defined holomorphic $(3,0)$ form (which is  a section of $K_X\otimes\call_\cald$) and $\varphi^{{\bf v}}$ and $\varphi^{\bar{\bf v}}$ are sections of the holomorphic normal bundle  $N_{S/X}$ and its complex conjugated $\bar N_{S/X}$. Furthermore, the chiral defects correspond to the intersections between the E3-brane and the bulk 7-branes characterizing these backgrounds and the two-dimensional chiral theories living thereon can be locally described in terms of open strings connecting the E3-brane to the bulk 7-branes. 

The papers \cite{Bianchi:2011qh,Bianchi:2012kt} focused on the four-dimensional fermionic sector, without explicitly considering the bosonic sector, the chiral two-dimensional sector  as well as the supersymmetric structure of the complete effective theory. These additional ingredients are described by our duality twisted action under certain  simplifying assumptions. In particular, it does not take into account bulk ingredients like axionic moduli,  fluxes and warping.  

In this section we are going to discuss how to incorporate the dependence on bulk axionic moduli. However notice that, even if we `turn off' all such bulk effects, the action (\ref{dtwaction}) would  not still  completely capture the physics of E3-branes for the following reason.  In general, the vacuum expectation value of the world-volume field-strength $F$ can be non-vanishing. More precisely, it preserves supersymmetry if and only if $F^{0,2}=F^{2,0}=0$ and $j^{mn}F_{mn}=0$ or, equivalently,
 \be 
 F=*F 
 \ee
 This condition can be derived from the twisted supersymmetry transformations (\ref{dtwsuper}) or, more generically,  from a standard  supersymmetry analysis for a probe E3-brane. The key point is that the action (\ref{dtwaction}) corresponds to the expansion  of the standard Dirac-Born-Infeld (DBI) effective action up to quadratic order in the field-strength $F$ around a  vacuum configuration in which $F$ is {\em vanishing}. In fact, when the background $F$ is non-vanishing, the action (\ref{dtwaction}) is not the whole story and in particular it misses some terms induced by the background $F$ which originate from the non-trivial DBI structure. Such terms can have important consequences. For example, they can induce a lifting  of zero-modes   \cite{Koerber:2006hh,Bianchi:2012pn,Bianchi:2012kt}. 
 
  However, in the present paper we ignore these important effects and proceed by  incorporating the non-trivial bulk ingredients in  (\ref{dtwaction}), obtaining a `simplified' E3-brane effective action and postponing the discussion of the complete E3-brane effective action to the future. 
From the M-theory viewpoint on F-theory compactification, this means that we are going to restrict to the standard (2,0) six-dimensional theory of the 
dual M5-brane instanton, neglecting additional terms which would be induced by the complete Born-Infeld-like Lagrangian, as for instance the one proposed in \cite{Bandos:1997ui}.

 \subsection{Adding axionic moduli}
 
Let us discuss the coupling to the various kinds of bulk potentials.  Type IIB supergravity contains the NS-NS $B_{\it 2}$, and the R-R $C_{\it 2}$ and  $C_{\it 4}$. The pair $(C_{\it 2},B_{\it 2})$ transforms as a doublet under the ${\rm SL}(2,\mathbb{Z})$ duality group.   By using this property, one can construct the 
${\rm SL}(2,\mathbb{Z})$ invariant four-form $\tilde C_{\it 4}\equiv C_{\it 4}-\frac12 B_{\it 2}\wedge C_{\it 2}$. In this paper we restrict to the fluxless case. Hence the corresponding field-strengths must be vanishing. 

In addition, there are gauge fields supported on the 7-branes characterizing the F-theory background. Suppose that there  are just single, possibly intersecting, $[p,q]$ 7-branes. For each of them one can then go to a local frame in which it corresponds to a D7-brane. It then  supports a world-volume gauge field $\hat A$. The associated field-strength $\hat F$ naturally combines with $B_{\it 2}$  into $\hat\calf=\frac{1}{2\pi}\, \hat F-B_{\it 2}|_{\rm D7}$.
The gauge field $\hat A$ transforms in such a way that $\hat\calf$ is gauge invariant under the gauge transformations of $B_{\it 2}$.
In fact $\hat\calf$ is a gauge-invariant source for the other bulk fluxes and then we set it to zero: $\hat\calf=0$.


The E3-brane couples directly to $\tilde C_{\it 4}$  and to the bulk K\"ahler moduli  by  the topological term
\be\label{topterm}
I_{\rm top}=-\pi\int_S j\wedge j-2\pi\ii \int_S\tilde C_{\it 4}
\ee
whose real part is proportional to the volume of $S$.\footnote{In other conventions $*j=-j$ and then ${\rm vol}_4=-\frac12j\wedge j$.} On the other hand, the two-form potentials naturally mix with the world-volume field-strength. In particular,  $B_{\it 2}$   combines with the world-volume gauge field in the field-strength $\calf=\frac{1}{2\pi}\, F-B_{\it 2}|_S$,
which is gauge invariant under the $B_{\it 2}$ gauge transformations. The YM action $I_{\rm YM}$ in (\ref{dtwaction}) must then be generalized to
\be\label{E3YM}
\begin{aligned}
I^{\rm E3}_{\rm YM}=&-\frac{\ii}{4\pi}\int_S\tau F\wedge F- 2\pi\int_S\Im\tau \calf_-\wedge \calf_-\cr
&-\ii\int_S(C_{\it 2}-\tau B_{\it 2})\wedge F+\pi\ii\int_S( B_{\it 2}\wedge C_{\it 2}-\tau B_{\it 2}\wedge B_{\it 2})
\end{aligned}
\ee
 where $\calf_\pm=\frac12(1\pm*)\calf$. $I^{\rm E3}_{\rm YM}$ indeed reduces to $I_{\rm YM}$  in (\ref{dtwaction}) by setting $B_{\it 2}= C_{\it 2}=0$.\footnote{Notice that, as $\tau$, the fields $B_{\it 2}$ and $C_{\it 2}$ are not globally defined but can jump at the  duality walls.} On the other hand,  $I_B$ and $I_F$ in (\ref{dtwaction}) are not modified by the presence of the gauge potentials. It is important to stress that $\sqrt{\Im\tau}\,\calf_\pm$ transform with ${\rm U(1)}_{\cald}$ charges $q_\cald=\pm 1$ under ${\rm SL}(2,\mathbb{Z})$ duality transformations, generalizing (\ref{sdfluxes}).

The three-dimensional contributions supported on the duality walls are still given by the CS-like terms described in section \ref{sec:dw}. In particular,  for the generating $T$ and $S$ dualities they are given by (\ref{csT}) and (\ref{csS}) respectively.

The two-dimensional theories supported on the chiral defects, that is, at the intersection of the E3-brane and the bulk 7-branes, naturally couple to the seven-brane gauge potentials. By adopting the local description in terms of $[1,0]$ chiral defects, they sit on the holomorphic two-cycle $\calc$ given by the intersection of the E3-brane and a bulk D7-brane. It is well-known that the chiral theory is given by a chiral fermion/boson with charges $(1,-1)$ under ${\rm U(1)_{\rm E3}\times  U(1)_{\rm D7}}$, which means that we must substitute $A$ with the combination $A-\hat A$ in the chiral theory. Furthermore, as it will be presently clear, one must also add the term $-\frac{\ii}{4\pi}\int_\calc A\wedge \hat A$. Hence bosonic chiral theory described by (\ref{2dB}) must be generalized to
\be\label{2dE3}
I^{\rm E3}_{\rm 2d}=\frac{1}{8\pi}\int_\calc *(\d\beta-A+\hat A)\wedge(\d\beta-A+\hat A)+\frac{\ii}{4\pi}\int_\calc\beta (F-\hat F)-\frac{\ii}{4\pi}\int_{\calc}A\wedge \hat A
\ee
One can easily check that,  under a gauge transformation $A\rightarrow A+\d\lambda$, $I^{\rm E3}_{\rm 2d}$  produces an anomalous   contribution $\Delta I^{\rm E3}_{\rm 2d,B}=\frac{\ii }{4\pi}\int_\calc\lambda F$ which is perfectly cancelled by the variation of the $T^{-1}$ wall contribution (\ref{3danomaly}).

In summary, the four-dimensional action (\ref{dtwaction}) must be generalized to
\be\label{E34d}
I^{\rm E3}_{\rm 4d}=I_{\rm top}+I^{\rm E3}_{\rm YM}+I_B+I_F
\ee
and must be supplemented by  (\ref{bb}) and   (\ref{2dE3}) (in presence of just $[1,0]$-chiral defects) into the complete effective action
\be\label{E3eff}
I_{\rm E3}=I^{\rm E3}_{\rm 4d}+ I_{\rm 3d}  +I^{\rm E3}_{\rm 2d}
\ee

\subsection{Supersymmetry}

Let us now check that (\ref{E3eff}) is invariant under the following natural modification of the duality twisted  supersymmetries (\ref{dtwsuper}):
\be\label{dtwsuper2}
\begin{aligned}
\tilde\calq_{\dot\alpha}\, A^{1,0}&=\frac{1}{\sqrt{\Im\tau}}\, \tilde\psi_{\dot\alpha} \quad~~~~~~~~~~~~~~~~~  \tilde\calq_{\dot\alpha}\, A^{0,1}=0\\
\tilde\calq_{\dot\alpha}\,  \sigma&=0\quad~~~~~~~~~~~~~~~~~~~~~~~~~~~~  \tilde\calq_{\dot\alpha}\,  \tilde\sigma=\tilde\rho_{\dot\alpha}\\
\tilde\calq_{\dot\alpha}\,  \varphi^{\beta\dot\gamma}&=\delta^{\dot\gamma}_{\dot\alpha}\lambda^\beta \quad~~~~~~~~~~~~~~~~~~~~~~~~
 \tilde\calq_{\dot\alpha}\, \tilde\lambda^{\dot\beta}=-\frac12\pi \sqrt{\Im\tau}\,\delta^{\dot\beta}_{\dot\alpha}\,j^{mn}\,\calf_{mn}\\
\tilde\calq_{\dot \alpha}\, \lambda^\beta&=0\quad~~~~~~~~~~~~~~~~~~~~~~~~~~~~~
\tilde\calq_{\dot\alpha}\, \tilde\psi^{\dot\beta}=\frac{\ii}{2}\delta^{\dot\beta}_{\dot\alpha}\del_\cala^\dagger\sigma\\
\tilde\calq_{\dot\alpha}\, \psi_\beta&=-\frac{\ii}{2}\delbar\varphi_{\beta\dot\alpha}\quad~~~~~~~~~~~~~~~~~~~ \tilde\calq_{\dot\alpha}\, \tilde\rho^{\dot\beta}=-2\pi\ii\sqrt{\Im\tau}\, \delta^{\dot\beta}_{\dot\alpha}\, \calf^{0,2}\\
\tilde\calq_{\dot\alpha}\, \rho^{\beta}&=0\quad~~~~~~~~~~~~~~~~~~~~~~~~~~~~~ \tilde\calq_{\dot\alpha}\, \beta=0
\end{aligned}
\ee
In passing, notice that a bosonic configuration with non-trivial $\calf$  is supersymmetric only if  $\calf^{0,2}=\calf^{2,0}=0$ and $j^{mn}\calf_{mn}=0$, i.e. if $\calf$ is self-dual
\be
*\calf=\calf
\ee
which indeed reproduces the supersymmetry condition obtained from standard $\kappa$-symmetry arguments for a probe E3-brane, see e.g.\ \cite{Bianchi:2011qh,Bianchi:2012kt}.

One can straightforwardly  check that $\delta I_{\rm 4d}^{\rm E3}$, with $\delta=\zeta_{\dot\alpha}\tilde\calq^{\dot\alpha}$, vanishes up to boundary terms. Hence, we have to re-examine
the contributions of duality walls and chiral defects. 
We  first consider the neighbourhood of  the intersection of the E3-brane and a D7-brane, which has the structure given in figure \ref{brc}.
In order to proceed, we have first to comment on a small  subtlety. In absence of D7-branes, one can associate with  $C_{\it 2}$  the  field-strength   $\tilde G_{\it 3}\equiv \d (C_{\it 2}-C_{\it 0}B_{\it 2})$. Notice that it is closed but  not gauge invariant under gauge transformations of $B_{\it 2}$, the gauge invariant field-strength being given by $G_{\it 3}=\tilde G_{\it 3}+B_{\it 2}\wedge F_{\it 1}$. 
On the other hand, in presence of    a non-trivial world-volume field $\hat F$ on the D7-brane, $\tilde G_{\it 3}$  is not everywhere closed anymore, rather it should satisfy the Bianchi identity $\d \tilde G_{\it 3}=-\frac{1}{2\pi} \hat F\wedge \delta^{\it 2}({\rm D7})$. One can then  locally write $\d [\tilde G_{\it 3}+\frac{1}{2\pi} \hat A\wedge \delta^{\it 2}({\rm D7})]=0$ and locally define $C_{\it 2}$ through  $\d (C_{\it 2}-C_{\it 0}B_{\it 2})=\tilde G_{\it 3}+\frac{1}{2\pi} \hat A\wedge \delta^{\it 2}({\rm D7})$. Hence, for vanishing gauge invariant flux $G_{\it 3}=0$ we see that\footnote{Notice that $C_{\it 2}-C_{\it 0}B_{\it 2}$ is invariant under a $T$ transformation. Hence, its does not `jump' when one crosses a $T$ wall and its exterior derivative does not contain any contribution localized on the wall. Hence, in the following formulas we write $\d\tau=G_{\it 1}+\ii\d e^{-\phi}$, without any delta-like contribution on the $T$ wall.} 
\be\label{modC2}
\d (C_{\it 2}-\tau B_{\it 2})=-\d\tau\wedge B_{\it 2}+\frac{1}{2\pi} \hat A\wedge \delta_X^{\it 2}({\rm D7})
\ee

Now, by using (\ref{modC2}) one  obtains that the variation of $I_{\rm 4d}^{\rm E3}$  under $\delta\equiv \zeta_{\dot\alpha}\tilde\calq^{\dot\alpha}$ gives  \footnote{In our conventions, the complex coordinates on $X$ can be written as $z^i=y^i+\ii y^{3+i}$, $i=1,2,3$, in terms of an oriented set of real coordinates $y^1,\ldots, y^6$. This implies that if $D_1$ and $D_2$ are two holomorphic four-dimensional submanifolds of $X$ and $\calc=D_1\cap D_2$, then 
$\delta^{\it 2}_X(D_1)\wedge \delta^{\it 2}_X(D_2)=-\delta^{\it 4}_X(\calc)$. } 
\be
\delta I^{\rm E3}_{\rm 4d}=-\frac{1}{2\pi\ii}\int_\calb\frac{1}{\sqrt{\Im\tau}}\,\zeta_{\dot\alpha}\tilde\psi^{\dot\alpha}\wedge F-\frac{1}{2\pi\ii}\int_\calc \frac{1}{\sqrt{\Im\tau}}\zeta_{\dot\alpha}\tilde\psi^{\dot\alpha}\wedge\hat A
\ee
On the other hand, the variation of three-dimensional term (\ref{cschiral}) produces 
\be
\delta I_{\rm 3d}=\frac{1}{2\pi\ii}\int_\calb\frac{1}{\sqrt{\Im\tau}}\,\zeta_{\dot\alpha}\tilde\psi^{\dot\alpha}\wedge F+\frac{1}{4\pi\ii}\int_\calc\frac{1}{\sqrt{\Im\tau}}\,\zeta_{\dot\alpha}\tilde\psi^{\dot\alpha}\wedge A
\ee
while the variation of the two-dimensional action (\ref{2dE3}) is
\be
\delta I^{\rm E3}_{\rm 2d}=\frac{1}{4\pi\ii}\int_\calc \frac{1}{\sqrt{\Im\tau}}\zeta_{\dot\alpha}\tilde\psi^{\dot\alpha}\wedge (2\hat A- A)
\ee
 We see that $\delta I_{\rm E3}=\delta I^{\rm E3}_{\rm 4d}+\delta I_{\rm 3d}+\delta I^{\rm E3}_{\rm 2d}=0$ and then the complete action is locally invariant
  by a delicate interplay between the four- three- and two-dimensional contributions. 
 
This discussion applies locally around the intersections of the E3-brane with any (isolated)  7-brane, by properly adjusting the duality wall network as in figure \ref{res}.  Hence, it remains to check  that more general duality walls are not dangerous for supersymmetry. This can be done as in section \ref{sec:susy}, taking into account that the combination $\frac{1}{\sqrt{\Im\tau}}(C_{\it 2}-\tau B_{\it 2})$ transforms as a two-form taking values in $\call^{-1}_\cald$ under ${\rm SL}(2,\mathbb{Z})$.

\subsection{Bulk gauge transformations}

  One can also check the invariance of the above E3-brane effective action under gauge transformations of the bulk potentials.
We just briefly discuss the case of a U(1)-gauge transformation $\hat A\rightarrow \hat A+\d\hat\lambda$ of the D7-brane world-volume. Taking into account that $\beta\rightarrow\beta-\hat\lambda|_\calc$, the variation of the two-dimensional action (\ref{2dE3}) produces the anomalous contribution
\be
\frac{\ii}{4\pi}\int_\calc\hat\lambda(\hat F-2  F)
\ee
 This is cancelled by an inflow mechanism from the four-dimensional action, by taking into account that  $C_{\it 2}$ and $\tilde C_{\it 4}$ transform as follows: $C_{\it 2}\rightarrow C_{\it 2}+\frac{1}{2\pi}\hat\lambda\delta_X^{\it 2}({\rm D7})$ and $\tilde C_{\it 4}\rightarrow \tilde C_{\it 4}-\frac{1}{4\pi}\hat\lambda\hat\calf\wedge\delta_X^{\it 2}({\rm D7})$.  The first can be easily obtained from (\ref{modC2}). The second can be analogously derived by solving  the Bianchi identity which defines $\tilde C_{\it 4}$ in presence of a D7-brane.\footnote{In presence of a D7-brane, the (non gauge-invariant) closed five-form field-strength $\tilde G_{\it 5}$ satisfies the Bianchi identity $\d  \tilde G_{\it 5}=\frac{1}{8\pi^2}\hat F\wedge \hat F\wedge \delta_X^{\it 2}({\rm D7})$. Away from the D7-branes, it is locally related to $C_{\it 4}$ by $ \tilde G_{\it 5}=\d(C_{\it 4}-B_{\it 2}\wedge C_{\it 2}+\frac12C_{\it 0}B_{\it 2}\wedge B_{\it 2})$. Hence, if $\hat F\neq 0$, this relation must be generalized to $\tilde G_{\it 5}-\frac{1}{8\pi^2}\hat A\wedge \hat F\wedge \delta_X^{\it 2}({\rm D7})=\d(C_{\it 4}-B_{\it 2}\wedge C_{\it 2}+\frac12 C_{\it 0}B_{\it 2}\wedge B_{\it 2})=\d(\tilde C_{\it 4}-\frac12B_{\it 2}\wedge C_{\it 2}+\frac12C_{\it 0}B_{\it 2}\wedge B_{\it 2})$. By using $C_{\it 2}\rightarrow C_{\it 2}+\frac{1}{2\pi}\hat\lambda\delta_X^{\it 2}({\rm D7})$, this implies that  $\tilde C_{\it 4}\rightarrow \tilde C_{\it 4}-\frac{1}{4\pi}\hat\lambda\hat\calf\wedge\delta_X^{\it 2}({\rm D7})$. } This cancellation is indeed a particular manifestation of the inflow mechanism discussed in \cite{Green:1996dd,Cheung:1997az,Minasian:1997mm}.


\section{An application}
\label{sec:application}

The partition function of our duality twisted theory, with appropriate insertions of operators dictated by the presence of zero-modes,  should provide information about the  F-terms in the effective four-dimensional theory of the corresponding F-theory compactifications  \cite{Becker:1995kb,Witten:1996bn,Harvey:1999as,Witten:1999eg,Beasley:2003fx,Beasley:2005iu}. In particular,  a superpotential is produced if there are no other fermionic zero modes in addition to the   two universal ones, that are the goldstinos  associated with the breaking by the brane instanton of two of the four bulk supersymmetries. From the fermionic action $I_{\rm F}$ in (\ref{dtwaction}) it is easy to identify the explicit structure of the fermionic zero-modes, see \cite{Bianchi:2011qh,Bianchi:2012kt} for a detailed discussion. The fermionic zero modes can be identified with the cohomology classes in $H^i(S,\calo_S)$ and $H^{i}(S,\call^{-1}_\cald)$, with $i=0,1,2$. In particular the two universal zero modes $\theta^\alpha$ correspond to $H^0(S,\calo_S)$ (they are given by  constant $\lambda^\alpha$) and then  all other cohomology groups must vanish in order for the superpotential not to be  identically vanishing.

The corresponding contribution  to the effective action is just provided by the partition function of the E3-brane effective theory
\be\label{pathint}
 \int\d^4 x\d^2\theta\, W_{\rm np}=\int\cald X\, e^{-I_{\rm E3}}
\ee
 where  $\cald X$ denotes the collective path-integral integration measure. 
 On the l.h.s.\ of (\ref{pathint}) we see the appearance of  the  universal  fermionic and bosonic  zero-modes   $\theta^\alpha$ and $x^\mu$,  the latter corresponding to constant  $\varphi^{\alpha\dot\beta}\sim\varphi^\mu$, which are  singled out from $\cald X$.  In this paper we are not going to discuss the important issues related to the explicit evaluation of the superpotential and other F-therms, postponing them to future work together with a more careful treatment of  anomalies and  other quantum aspects. Nevertheless, one can still try to run various formal arguments used in more standard  topological theories, see e.g.~\cite{Witten:1988ze,Witten:1994ev,Vafa:1994tf}. Here we are just going to give a simple example. 
 
 By  holomorphy arguments \cite{Witten:1996bn} based on the low-energy effective theory of the F-theory compactification, it is expected that  the superpotential  (\ref{pathint})  depends on the background K\"ahler form $J$ just  through the topological $I_{\rm top}$ contribution (\ref{topterm}) to $I_{\rm E3}$. This statement has never been explicitly derived at some direct microscopic level. We are now going to use our duality twisted theory to provide such a derivation.

Let us split the E3-brane action (\ref{E3eff}) as 
\be
I_{\rm E3}=I_{\rm top}+I
\ee
$I_{\rm top}$ clearly depends on $J$ through its pull-back $j$. Such a dependence is cohomological in nature and indeed  the factor $e^{-I_{\rm top}}$ in (\ref{pathint}) gives a well defined dependence of the four-dimensional effective superpotential on the K\"ahler moduli. The statement we want to prove  is that the dependence on the K\"ahler moduli
of $ W_{\rm np}$ is exhausted by this $e^{-I_{\rm top}}$ factor. Namely, we are going to show (at a formal level) that the partition function
\be
Z=\int\cald X'\, e^{-I}
\ee
is independent on the background K\"ahler form $J$. In the measure $\cald X'$ we have omitted the universal zero-mode factor $\d^4 x\d^2\theta$.

Consider a general deformation $\delta J$ of the background K\"ahler form. $\delta J$ is given by a general (small) closed $(1,1)$ form.  This induces a deformation $\delta j$ of the world-volume K\"ahler form. We then need to prove that $Z$ is invariant under such deformation. 
  The strategy is quite standard in the context of topological theories.
  Namely, the key step is to prove that the variation  of the action under  $\delta j$   can be written as a $\tilde\calq_{\dot\alpha}$-exact term:
  \be\label{exact}
  \delta I=\int_{S}\{\tilde\calq_{\dot\alpha},\delta\calv^{\dot\alpha}\}
  \ee
 where $\delta\calv^{\dot\alpha}$ is a globally defined fermionic operator. We  have used operatorial notation in which the 
variation of any operator $\calo$ under $\tilde\calq_{\dot\alpha}$ is denoted by $\{\tilde\calq_{\dot\alpha},\calo\}$.

 In order to prove (\ref{exact}), it is convenient to go to an equivalent formulation with auxiliary fields\footnote{I would like to thank S.~Giusto for discussions on this point.}, which is obtained by substituting $I_{\rm B}$ in   (\ref{dtwaction}) with 
\be
 I'_{\rm B}=-\frac{\ii}{4\pi}\int_S j\wedge \del\varphi^{\alpha\dot\beta}\wedge \delbar\varphi_{\alpha\dot\beta}  -\frac{\ii}{2\pi}\int_S j\wedge h\wedge\tilde h+\frac{1}{2\pi}\int_S \big(h\wedge \del_\cala\tilde\sigma+\tilde h\wedge\delbar_\cala\sigma\big)
\ee
This action contains two new auxiliary fields    $h$ and $\tilde h$, taking values in $\Lambda^{1,0}\otimes \call_\cald$ and   $\Lambda^{0,1}\otimes \call^{-1}_\cald$ respectively.
One can readly verify that  by integrating out $h$ and $\tilde h$ one gets back $I_{\rm B}$. 
The twisted supersymmetry transformations (\ref{dtwsuper2}) must be modified accordingly, by making the substitution  
\be\label{modsusy}
\tilde\calq_{\dot\alpha}\,\tilde\psi^{\dot\beta}=\frac{\ii}{2}\delta^{\dot\beta}_{\dot\alpha}\del_\cala^\dagger\sigma\quad\rightarrow \quad\tilde\calq_{\dot\alpha}\,\tilde\psi^{\dot\beta}=\frac{\ii}{2}\delta^{\dot\beta}_{\dot\alpha}h\,,\quad\tilde\calq_{\dot\alpha}\, h=0\,,\quad\tilde\calq_{\dot\alpha}\, \tilde h=-\delbar_\cala\tilde\lambda_{\dot\alpha}
\ee
The formulation with auxiliary fields $h$ and $\tilde h$ is clearly advantageous for the present problem since it has a simpler dependence on the  \kahler form $j$.   

One can now compute the variation of $I$ under  $\delta j$, the only slightly tricky part being given by the variation of the $\int_S\Im\tau \calf_-\wedge \calf_-$ part in $I^{\rm E3}_{\rm YM}$. The result is
\be
\begin{aligned}
\delta I=&-\pi\int_S\Im\tau\,(j^{mn}\calf_{mn})\, \calf\wedge \delta j_{\rm P}-\frac{\ii}{4\pi}\int_S \delta j\wedge \del\varphi^{\alpha\dot\beta}\wedge \delbar\varphi_{\alpha\dot\beta}  -\frac{\ii}{2\pi}\int_S \delta j\wedge h\wedge\tilde h\\
&+\frac{1}{\pi}\int_S\big(\delta j\wedge \del\lambda^\alpha\wedge\psi_\alpha+\delta j\wedge\tilde\psi_{\dot\alpha}\wedge \delbar_\cala\tilde\lambda^{\dot\alpha}\big)
\end{aligned}
\ee 
Notice that, in the first term on the right-hand side $\delta j$ appears only through its primitive component $\delta j_{\rm P}\equiv \delta j-\frac14 (j^{mn}\delta j_{mn})j$. Then, from the supersymmetry  transformations (\ref{dtwsuper2}) with the substitution (\ref{modsusy}), it is possible to rewrite $\delta I$ as in (\ref{exact}) with\footnote{\label{foot:contact}More precisely, with the choice (\ref{calv}), equation  (\ref{exact}) is valid by neglecting a composite operator  $\tilde\lambda_{\dot\alpha}\delbar_\cala \tilde\psi^{\dot\alpha}\wedge j$ in $\delta I$. This clearly  vanishes at the classical level  by the equation of motion $ \delbar_\cala \tilde\psi_{\dot\alpha}\wedge  j=0$  and  we are implicitly  assuming that we are using a regularization scheme in which this statement is preserved  at the quantum level, see for instance \cite{Collins:1984xc}. In any case, in presence of insertions, such term could generate additional contact terms.} 
\be\label{calv}
\delta\calv^{\dot\alpha}=\sqrt{\Im\tau}\, \tilde\lambda^{\dot\alpha}\,\calf\wedge \delta j_{\rm P}-\frac{1}{2\pi}\psi_{\beta}\wedge\del\varphi^{\beta\dot\alpha}\wedge\delta j-\frac1{2\pi}\,\tilde\psi^{\dot\alpha}\wedge\tilde h\wedge\delta j
\ee
Importantly, only the primitive $(1,1)$ component of $\sqrt{\Im\tau}\calf$ enters in $\delta\calv^{\dot\alpha}$. This transforms as a section of $\call_\cald$ under ${\rm SL}(2,\mathbb{Z})$. All the other fields in  $\delta\calv^{\dot\alpha}$ transform as specific powers of  $\call_\cald$ too  and one can easily check that the different terms in $\delta\calv^{\dot\alpha}$ are actually invariant under ${\rm SL}(2,\mathbb{Z})$. Hence, $\calv^{\dot\alpha}$ as well as $\{\tilde\calq_{\dot\alpha},\calv^{\dot\alpha}\}$  are  {\em invariant} under the ${\rm SL}(2,\mathbb{Z})$ duality.

Equation (\ref{exact}) is all one needs to run a standard argument. Indeed, we formally  get
  \be\label{varZ}
  \delta Z= -\int\cald X' \,\delta I \, e^{-I}  = -\int\cald X' \,\int_{S}\{\tilde\calq_{\dot\alpha},\calv^{\dot\alpha}\}\, e^{-I} =-\int_{S}\langle\{\tilde\calq_{\dot\alpha},\calv^{\dot\alpha}\}\rangle
\ee
Since for a supersymmetric theory $\langle\{\tilde\calq_{\dot\alpha},\calo\}\rangle=0$ for any operator $\calo$, one obtains the desired result
\be
  \delta Z=0
\ee
The fact that, as stressed above, $\{\tilde\calq_{\dot\alpha},\calv^{\dot\alpha}\}$ is invariant under the ${\rm SL}(2,\mathbb{Z})$ duality group  is an important prerequisite for the self-consistency of the argument, since only in this case  it is sensible to integrate $\{\tilde\calq_{\dot\alpha},\calv^{\dot\alpha}\}$ over $S$, getting a well defined  right-hand side of (\ref{varZ}). This provides a non-trivial consistency check of the our approach.

As usual, this argument can be extended to correlation functions of operators that are invariant under $\tilde\calq_{\dot\alpha}$.  Indeed, if $\calo$ is some bosonic $\tilde\calq_{\dot\alpha}$-invariant operator such that $\{\tilde\calq_{\dot\alpha},\calo\}=0$, then\footnote{Here we are assuming that the compact terms discussed footnote \ref{foot:contact} are not relevant.} 
\be
\delta   \langle\calo \rangle=\int\cald X (\delta\calo-\calo \,\delta I)  \,e^{-I}  = \langle\delta\calo \rangle -\int_{S}\langle\{\tilde\calq_{\dot\alpha},\calo\calv^{\dot\alpha}\}\rangle =\langle\delta\calo \rangle 
\ee
where $\delta\calo$ gives the variation of $\calo$ under the deformation of the K\"ahler form. In particular, if $\calo$ does not explicitly depend on $j$, then
\be
\delta   \langle\calo \rangle=0
\ee

\section{Conclusions}
\label{sec:concl}

In this paper we have introduced the idea of topological duality twist of $N=4$ SYM theories, which extends the standard topological twist to involve ${\rm SL}(2,\mathbb{Z})$ duality transformations. We have focused on the case in which the gauge group is just U(1), the space is  \kahler    and the coupling $\tau$ varies holomorphically on it, which is the relevant one for studying D3-brane instantons in F-theory compactifications. This paper provides just a first step in the exploration of such theories and there are still several important open issues which remain to be addressed. 

At a more theoretical level, it is important to go beyond the basically classical approach used here, investigating quantization, observables and anomalies more in detail. 
As we have already checked for some gauge anomalies, since D3-brane instantons provide a stringy (hence, non-anomalous) realization of the system, we expect anomaly cancellations mechanisms to be available, as for instance in \cite{Bachas:1999um}. This should involve the inclusion of curvature terms \cite{Green:1996dd,Cheung:1997az,Minasian:1997mm}, a possible refined interpretation of the world-volume gauge field as in \cite{oai:arXiv.org:hep-th/9907189}
and the identification of the appropriate spin structures associated with the two-dimensional chiral theories \cite{Witten:1996hc}.  Furthermore,  it would be very interesting to study the possible realization of the  topological duality twist for  higher rank gauge groups. 
This extension would be presumably connected with results of \cite{Gaiotto:2008ak} and would enrich significantly the physical content of the theory.

At a more applicative level, we hope that the present work could provide a concrete starting point 
for the development of technical tools in the computation of non-perturbative corrections to the effective theory of F-theory compactifications. 
In this respect, the results of the present work should be combined with those of \cite{Bianchi:2012pn,Bianchi:2012kt} regarding the structure of fermionic zero modes.
In particular, one should incorporate the effect of the massive terms induced by bulk and world-volume fluxes which originates  from the complete Dirac-Born-Infeld D3-brane effective action and are not captured by the duality twist of the $N=4$ SYM theory.


\vspace{1cm}

\centerline{\large\em Acknowledgments}

\vspace{0.5cm}

\noindent I would like to thank M.~Bianchi, A.~Collinucci, S.~Giusto, A.~Grassi, C.~Imbimbo, G.~Inverso, K.~Lechner, P.A.~Marchetti, D.R.~Morrison, E.~Plauschinn and  R.~Valandro
for fruitful discussions. I especially thank I.~Garcia-Etxebarria, D.~Sorokin  and T.~Weigand  for  useful discussions and comments on the draft. This work is partially supported by MIUR-PRIN contract 2009-KHZKRX, by the Padova University Project CPDA119349 and by INFN.

\vspace{1cm}










\end{document}